\documentclass[astrosymb]{aastex631}

\usepackage{color}
\usepackage{graphicx}
\usepackage{dcolumn}
\usepackage{bm}
\usepackage{hyperref}
\usepackage{amsmath} 
\usepackage{soul}

\newcommand{\kpc}{\ensuremath{\rm{ kpc}}}
\newcommand{\msun}{\ensuremath{M_{\odot}}}

\def\apjl{{\rm ApJ}}                % Astrophysical Journal, Letters
\def\apjs{{\rm ApJS}}               % Astrophysical Journal, Supplement
\def\ao{{\rm Appl.~Opt.}}           % Applied Optics
\def\aap{{\rm A\&A}}                % Astronomy and Astrophysics
\begin{document}
\title{Extracting the Subhalo Mass Function from Strong Lens Images with Image Segmentation}
\shortauthors{Ostdiek, Diaz Rivero, and Dvorkin}
\shorttitle{Extracting the Subhalo Mass Function from Strong Lens Images with Image Segmentation}

\author[0000-0002-0376-6461]{Bryan Ostdiek}
\affiliation{
Department of Physics, Harvard University, Cambridge, MA 02138, USA
}
\email{bostdiek@g.harvard.edu}

\author[0000-0003-2123-049X]{Ana Diaz Rivero}
\affiliation{
Department of Physics, Harvard University, Cambridge, MA 02138, USA
}

\author[0000-0003-1476-1241]{Cora Dvorkin}
\affiliation{
Department of Physics, Harvard University, Cambridge, MA 02138, USA
}

\date{\today}
        
%%%%%%%%%%%%%%%%%%%%%%%%%%%%%%%%%%%%%%%%%%%%%%%%%%%%%%%%%%%%%%%%%%%%%%        
\begin{abstract}
Detecting substructure within strongly lensed images is a promising route to shed light on the nature of dark matter.
However, it is a challenging task, which traditionally requires detailed lens modeling and source reconstruction, taking weeks to analyze each system.
We use machine-learning to circumvent the need for lens and source modeling and develop a neural network to both locate subhalos in an image as well as determine their mass using the technique of image segmentation.
The network is trained on images with a single subhalo located near the Einstein ring across a wide range of apparent source magnitudes.
The network is then able to resolve subhalos with masses $m\gtrsim 10^{8.5}\msun$.
Training in this way allows the network to learn the gravitational lensing of light, and remarkably, it is then able to detect entire populations of substructure, even for locations further away from the Einstein ring than  those used in training.
Over a wide range of the apparent source magnitude, the false-positive rate is around three false subhalos per 100 images, coming mostly from the lightest detectable subhalo for that signal-to-noise ratio.
With good accuracy and a low false-positive rate, counting the number of pixels assigned to each subhalo class over multiple images allows for a measurement of the subhalo mass function (SMF).
When measured over three mass bins from $10^9\msun$--$10^{10} \msun$ the SMF slope is recovered with an error of 36\% for 50 images, and this improves to 10\% for 1000 images with Hubble Space Telescope-like noise.
\end{abstract}

% \tableofcontents
\section{Introduction}

The lambda cold dark matter ($\Lambda$CDM) paradigm has been successful at explaining many cosmological observations.
However, on smaller scales (galactic/sub-galactic), dark matter clustering depends on particulars of the dark matter model, so even if large-scale observations are consistent with a cold dark matter particle, probing small scales can reveal an exotic dark matter scenario. 
For instance, in warm dark matter~\citep{1994PhRvL..72...17D,2001ApJ...556...93B,2007PhRvD..76j3515H}, self-interacting dark matter~\citep{2018PhR...730....1T}, or ultra-light bosonic dark matter models~\citep{2000PhRvL..85.1158H,2017PhRvD..95d3541H}, overdensities below a certain threshold do not collapse to form bound structure, which creates a low-mass cutoff of the halo mass function.
For instance, warm dark matter with a mass of few keV can cut off the halo mass function around $10^7-10^8\msun$ with larger masses having lower cutoffs~\citep{2012MNRAS.424..684S,2016MNRAS.455..318B,2019arXiv191102663B,2020MNRAS.491.6077G}.
Therefore, searching for the low-mass dark matter halos serves as a test for the $\Lambda$CDM paradigm and can help reveal the nature of dark matter.

Unfortunately, low-mass halos are particularly hard to find.
At low masses $\left(m \lesssim 10^9\msun \right)$, star formation is suppressed and halos are not very luminous~\citep{1977MNRAS.179..541R,1992MNRAS.256P..43E,2010AdAst2010E...8K,2013RPPh...76k2901B,2017MNRAS.467.2019R,2017MNRAS.471.3547F}.
Even in the Local Group, it is challenging to detect light from these halos.
Gravitational interactions may then provide the best opportunity to detect subhalos that live within larger halos.
Local searches for dark matter subhalos include analyzing the effect of dark substructure passing through extended cold stellar streams~\citep{2014ApJ...788..181N,2016ApJ...820...45C,2016PhRvL.116l1301B,2016MNRAS.463..102E,2019ApJ...880...38B,2019arXiv191102663B}, careful examinations of stellar motions~\citep{2015MNRAS.446.1000F}, and combining collective motions of stars and stellar weak lensing~\citep{VanTilburg:2018ykj,Mondino:2020rkn,Mishra-Sharma:2020ynk}.

Outside of the Local Group, the only method to detect dark matter subhalos -- so far -- is strong gravitational lensing of galaxies or quasars.
In order to probe low-mass subhalos, the object acting as the lens needs be galaxy sized.
In galaxy--galaxy lensing, light from a distance source is deflected by a foreground galaxy.
Most of the mass, and therefore the shape of the lens, comes from the dark matter halo of the galaxy.
Dark matter substructure then serves as a perturbation to the main lens.
The amount of perturbation depends on the mass and location of the subhalo.

The traditional technique to detect substructure in strong lensing images as a first step involves modeling a smooth lens and the background source of light.
Residuals between an image generated by forward modeling through the best-fit smooth model and the data can then be minimized by adding substructure to improve the fit~\citep{1998MNRAS.295..587M, 2003MNRAS.339..607M, 2005MNRAS.363.1136K, 2009MNRAS.392..945V, 2013ApJ...767....9H}.
To date, there have been two systems found with evidence for resolved substructure, one with a mass of $\left(3.51 \pm 0.15 \right)\times10^9 \msun$~\citep{2010MNRAS.408.1969V} and one with a mass of $\left(1.9\pm 0.1\right)\times10^8\msun$~\citep{2012Natur.481..341V}.

The traditional methodology relies heavily on accurate modeling; inaccuracies can lead to extra residuals and false positives~\cite{2019MNRAS.485.2179R}.
The intensive modeling required for this technique makes the method slow and computationally expensive.
Additionally, alternative ways of modeling can lead to different results~\citep{2014MNRAS.442.2017V}.
For instance, both of the detected subhalos were originally modeled with pseduo-Jaffe profiles~\citep{2002ApJ...572...25D}.
However, the mass recovered when modeling instead with a Navaro-Frenk-White (NFW) profile is significantly higher (the $\left(3.51 \pm 0.15\right)\times10^9 \msun$ subhalo could be shifted to around $10^{10}\msun$~\citep{2018MNRAS.481.3661V}).
One extra caveat about these methods is that they are sensitive to the \emph{effective} mass of the subhalo.
\cite{2017ApJ...845..118M} showed that the true subhalo mass can be up to an order of magnitude larger than the effective subhalo that is reconstructed.

Other strategies to learn about the nature of dark matter from strong lenses bypass identifying and characterizing individual subhalos and instead look for the collective effect of a population of low-mass halos~\citep{2016JCAP...11..048H, 2016MNRAS.455.1819B, 2016PhRvD..94d3505C, 2017JCAP...05..037B, 2018ApJ...854..141D, 2018PhRvD..97b3001D, 2018PhRvD..98j3517D, 2019MNRAS.488.5085B}.\footnote{Statistical evidence for substructure was found in \cite{2017JCAP...05..037B}, allowing for a bound on the warm dark matter mass to be $m>2$ kev.}
Due to the steep low-mass end of the halo and SMFs in CDM, we expect an abundance of low-mass halos.
Although most of the individual halos are not detectable individually, their collective perturbations on an image can be detected.
This is desirable since dark matter theories make population-level predictions.
Another advantage of statistical methods is that they can be sensitive to lower masses (\emph{e.g.} \cite{2018PhRvD..98j3517D} showed that, on observable scales, the convergence power spectrum is most sensitive to the much more abundant population of $10^6 - 10^8 \msun$ halos than it is to the much smaller population of higher-mass halos).
Despite their advantages, these statistical methods still require either removing the main lens or simultaneously inferring both the mains lens and substructure.

As finding substructure has proved to be so difficult, machine-learning techniques are being applied to the problem.
Many uses have been found for these methods, from including better/faster modeling of the lens to measuring substructure directly.
\cite{2017Natur.548..555H} and \cite{2017ApJ...850L...7P} showed that a convolutional neural network (CNN) can be used to infer model parameters for the main lens.
With the inferred model parameters, one can then look for residuals
~\citep{2018arXiv180800011M, 2019ApJ...883...14M}.
Modeling the whole system (light source and main lens) was done in \cite{2019arXiv191006157C} using deep probabilistic programming combined with autoencoders.

Alternative machine-learning studies have aimed at looking for substructure directly.
A CNN was used for binary classification to determine if a strongly lensed image contains substructure beyond the main lens or not~\citep{2020PhRvD.101b3515D}.
It did not require individually modeling the main lens as a prior step to finding substructure, and allows for quickly identifying which images dedicated modeling should be focused on.
The lower mass reach for this method is around $10^8\msun$.
While this is closely related to the direct detection methods above, it is also possible to develop machine-learning methods to replicate the statistical searches previously discussed, meaning that they aim to uncover population-level characteristics without requiring individually resolving subhalos. 
\cite{2019ApJ...886...49B} built a network that assumes the presence of substructure and infers both the abundance of dark matter subhalos and the slope of the SMF.
Similarly, \cite{2020arXiv200505353V} use the effects of many subhalos to infer the low-mass cutoff of the SMF.
This is done as a classification problem, where the output for a given image is the mass bin where function is cutoff.
CNNs were trained on images with different types of dark matter substructure to learn how to distinguish them in \cite{2020ApJ...893...15A}.
\cite{NeurIPS2019_Lin} use a combination of DenseNet and ResNet to output a map corresponding to the probability that substructure exists at a specific location in an image.
As an alternative to these supervised learning methods, which require a labeled data set from which to train the network, \cite{2020arXiv200812731A} uses unsupervised techniques to detect subhalos in simulated images.

\begin{figure}[t]
\centering
\includegraphics[width=0.55\linewidth]{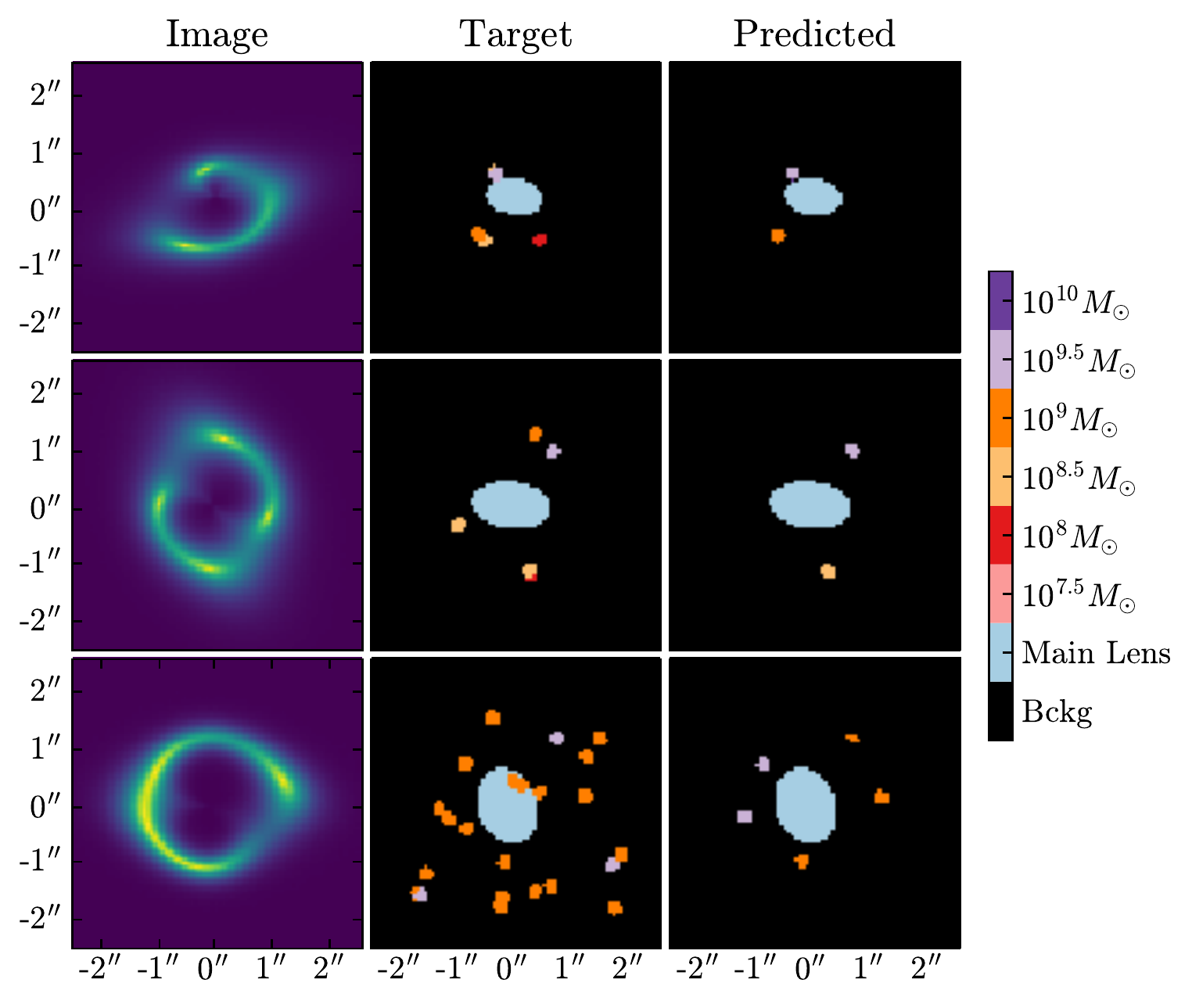}
\caption{
Examples of segmenting gravitationally lensed images. 
The left panels show simulated images that are fed into the neural network. 
Each pixel in an image is mapped to a label as either being the main lens within the Einstein radius, having a subhalo of a given mass, or as none of these (background). 
The middle panels show the true labels for the corresponding input image. 
The right panels show the corresponding output of our neural network.
The network was trained only on images with a single subhalo, but it is still able to find substructure in images with a rich population of subhalos.
}
\label{fig:Examples}
\end{figure}

In this work and our companion paper~\citep{Ostdiek:2020cqz}, we present results for a new method to directly detect substructure in strong gravitational lens images.
This is based on a machine-learning technique of object detection, where specific objects are found in an image.
This is done by training on many example images, which have been labeled by hand.
After the network has been trained, it is able to detect objects in images it has not seen before.

We focus on the particular technique of image segmentation, where rather than putting a box around the object, each pixel is classified as belonging to a specific class.
We use a U-Net~\citep{2015arXiv150504597R} architecture for this task, which consists of many convolutional layers, along with a down-sampling and up-sampling to help the network detect features at different scales.
The U-Net has emerged as one of the best architectures for image segmentation and was designed to track cells in biological images.\footnote{Architectures similar to the U-Net have been used in astrophysics/cosmology, \emph{e.g.} \cite{2020ApJS..248...20H} used it to predict galaxy morphologies.}
We use it to classify each pixel in an image as belonging to one of several predetermined classes the network is trained to identify.
Each pixel in our simulated images can fall into one of 11 different classes: part of the main lens, a subhalo with a mass within one of nine mass bins, or neither (background).
At the pixel level, this is a classification task.
However, it allows us to both locate and get the mass of substructure in the gravitational lens.

An example of our image segmentation is shown in Fig.~\ref{fig:Examples}.
The network takes in as input a lensed image as given by the left column and labels each pixel, quickly identifying both the main lens and the substructure, as shown by the right panel.
The mass of the subhalos is denoted by the color of the pixel.
In the middle column, we have used truth knowledge of the lens to label the area within the Einstein radius of the main lens and the different subhalos.
The network struggles to detect substructure near the edges of the image but is accurate for heavy subhalos near the Einstein ring.
We also see that it does not appear to add in spurious subhalos.
These two factors (good accuracy and low false-positive rate) allow us to apply this method to incorporate the step of combining images to put constraints on the SMF into a single analysis pipeline.

This paper is organized as follows. 
In Sec.~\ref{sec:data}, we discuss our image generation pipeline.
A detailed discussion of the network setup and the training regimen is contained in Sec.~\ref{sec:Model}.
We study the accuracy of the network and the false-positive rate in Sec.~\ref{sec:characterizing}.
In Sec.~\ref{sec:shmf}, we apply the network to images with multiple subhalos and infer the SMF.
We conclude in Sec.~\ref{sec:conclusion}.

%%%%%%%%%%%%%%%%%%%%%%%%%%%%%%%%%%%%%%%%%%%%%%%%%%%%%%%%%%%%%%%%%%%%%% 
\section{Data generation}
\label{sec:data}
%%%%%%%%%%%%%%%%%%%%%%%%%%%%%%%%%%%%%%%%%%%%%%%%%%%%%%%%%%%%%%%%%%%%%% 

The goal of our work is to detect dark substructure within strong lens images using the technique of image segmentation.
This is a supervised learning problem, so in order to train the network, a set of training data, including the target labels, is needed.
We generate strongly lensed images using the software package \textsc{Lenstronomy}~\citep{2018PDU....22..189B, 2015ApJ...813..102B}.
We use images with $80\times80$ pixels with a field of view of $5\arcsec\times5\arcsec$. This corresponds to a resolution of $0.06 \arcsec$ per pixel.
Each image contains a background source light, a smooth lens, sometimes substructure in the lens, noise, and it is always convolved with the default, pixel-based point spread function (PSF) of the Hubble Space Telescope (HST) WFC3\_F160W band in the \texttt{Lenstronomy} \texttt{SimulationAPI}.
In each image, the gravitational lens (main halo and subhalos) as well as the source light are unique.
Each step in the simulation pipeline is detailed below.

{\textbf{Smooth Lens}}: 
The halo of the main lens is modeled as a singular isothermal ellipsoid (SIE)~\citep{1994A&A...284..285K}.
In \textsc{Lenstronomy}, these lenses are parametrized by the Einstein radius ($\theta_E$) and the ellipticity moduli.
We choose the size of the Einstein radius to be typical of observed strongly lensed galaxy--galaxy systems, drawn from a uniform distribution
\begin{equation}
    \theta_E \in U[0.85, 1.50] \arcsec~,
\end{equation}
and an ellipticity drawn from
\begin{equation}
     q \in U[0.4, 1] ~, 
\end{equation}
where $q$ is the ratio of the minor-to-major axis of the ellipse.
The major axis can take on any angle in the image.
The center is chosen to be near the middle of the image so that the images/arcs lie within the field of view. The $x$ and $y$ positions are drawn randomly as
\begin{equation}
    x,y \in U[-0.25, 0.25] \arcsec~.
\end{equation}

To allow for more realistic lenses, we additionally add $m=3$ and $m=4$ multipole moments of the lensing potential to capture departures from ellipticity.
\textsc{Lenstronomy} uses the same definitions as in \cite{2013arXiv1307.4220X} 
The amplitude of each moment is chosen uniformly between
\begin{equation}
    a_m =\in U  [0, 0.1],
\end{equation}
and the orientation can take on any angle.

Finally, we allow for external shear to the lens.
We draw this uniformly in the range
\begin{equation}
    \gamma_{1,2}^{\text{shear}} \in U[-0.2, 0.2]~.
\end{equation}

In this work, we fix the distance to the lens at a redshift of $z_{\rm{lens}} = 0.2$.
Our fiducial cosmology is given by the Planck 2015 results in  \cite{Ade:2015xua}.  
These lens parameters (and the location of the source light) were chosen such that the main lens has a mass of order $10^{13}\msun$ (depending on the specific Einstein radius in a given image).

\begin{figure}[t]
    \centering
    \includegraphics[width=0.9\linewidth]{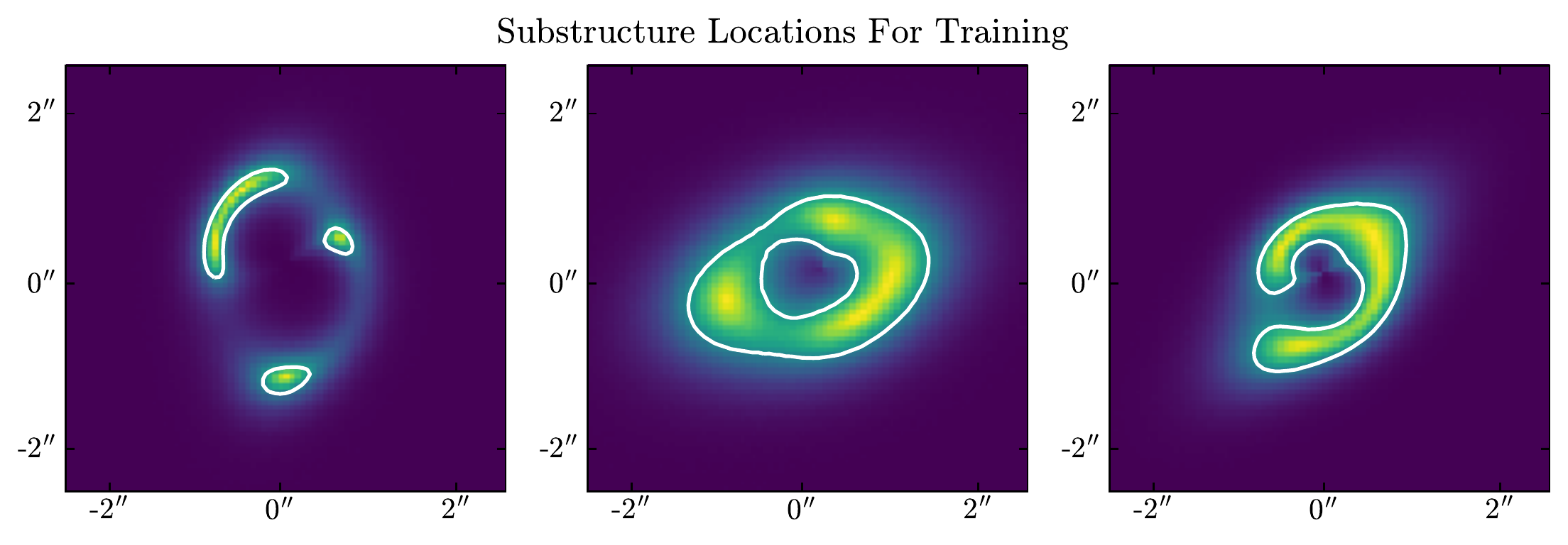}
    \caption{For training, the subhalos are restricted to areas where the image pixel brightness is at least 50\% as bright as the maximum pixel in the image.
    Examples of these regions are denoted by the white contours in the three panels shown.
    This definition does not fully follow the Einstein radius, but instead places the subhalos in areas where their effects may be large enough to be detected.
    }
    \label{fig:subahlo_locations}
\end{figure}

{\textbf{Subhalos:}}
When we add substructure to the lens, it is modeled as a truncated NFW profile~\citep{1996ApJ...462..563N,2009JCAP...01..015B} with a concentration parameter $c=15$.
The subhalos are truncated at five times the scale radius.
The network is trained on images that have either zero or one subhalo.
The subhalo masses are chosen to be log-uniform over the mass range $10^{5.75}-10^{10.25}\msun$, such that we obtain equal numbers of images in each mass bin.
For training, the subhalos are placed in bright pixels near the Einstein radius, where their effects are largest.
We define this as pixels that are at least $50\%$ as bright as the brightest pixel in the image.
Three examples of this are shown in Fig.~\ref{fig:subahlo_locations}.

In Sec.~\ref{sec:shmf}, we test a trained network on images with many subhalos.
For these images, the masses are drawn according to a power law given by
\begin{equation}
   \frac{dN}{dM} = a_0 \Bigg(\frac{M}{m_0} \Bigg)^{\beta},
   \label{eqn:aquarius}
\end{equation}
which was found to be a good fit of the subhalo population in the Aquarius simulation~\citep{2008MNRAS.391.1685S} with power law index $\beta=-1.9$, amplitude $a_0 = 3.26 \times 10^{-5} \msun^{-1}$, and pivot point of $m_0 = 2.52 \times10^7\msun$ although this specific normalization does not necessarily apply outside of the Aquarius simulation, the general form of Eq.~\eqref{eqn:aquarius} is universally found in $N$-body CDM simulations.
In such simulations, the three-dimensional distribution of subhalo positions is nearly spherically symmetric, with a strong dependence on the radius.
However, because their positions are projected onto a single plane, and the fact that strong lens images have a small field of view compared to the full extent of the halo perpendicular to the line of sight, a uniform distribution for subhalo positions is a good approximation (for example, see \cite{2018PhRvD..98j3517D}).
Thus, in our images with many subhalos, we will draw their locations uniformly across the whole image.

{\textbf{Target Labels:}}
The target labels, which the network is trying to predict, are generated only from the smooth lens and substructure, and do not use information from the source light or the observed image.
Pixels with a projected surface mass density larger than the critical density (convergence greater than 1) are labeled as the main lens class.
For each subhalo, we draw a circle with a radius of 2 pixels centered on its location and assign all the pixels within the circle as belonging to a given subhalo mass bin class.
Any pixel which has not been labeled as the main lens or a subhalo is denoted as background.
This method of identifying subhalos treats all subhalo masses as identical in that the more (less) massive subhalos do not get larger (smaller) circles, even though their effects are larger (smaller).
We choose to do this for two reasons.
The first is that it leads to more stable training of the network.
When the pixel labels change size with different masses, there are many more training pixels for the heavier classes than the light classes.
This creates an imbalance that would need to be corrected for, be it by weighting subhalo classes differently in the loss function or having different numbers of training images for each subhalo mass bin.
The second reason is that it makes counting subhalos easier.
The predicted subhalo count can be obtained by dividing the total number of pixels predicted to be part of a subhalo mass bin by the expected area per subhalo ($4\pi ~\rm{pixels}$).
Furthermore, as we will discuss extensively in Sec.~\ref{sec:shmf}, we are interested in extracting the SMF from an ensemble of images, for which we simply need the number of subhalos in each mass bin, and therefore do not need to faithfully reconstruct the surface mass density on the lens plane.
It would also be interesting to allow for the size of the predicted subhalos to vary according to the uncertainty in the localization. However, this is beyond the scope of this paper and left for future work.

{\textbf{Source Light:}}
This work is focused on images of lensed extended objects because they offer better chances for detecting substructure than point source-like objects~\citep{2013ApJ...767....9H}.
As a proof-of-principle, we use light sources modeled with an elliptical Sersic profile. 
The radius of the light source is in the range
\begin{equation}
    R_{\rm{ser}} \in U[0.1~\kpc, 0.8~\kpc] 
\end{equation}
and the ratio of the minor-to-major axes of the ellipse is
\begin{equation}
    q_{\rm{ser}} \in [0.33, 1.0]
\end{equation}
with the orientation randomly chosen as
\begin{equation}
    \theta_{\rm{ser}} \in [-\pi, \pi].
\end{equation}
The light is placed near the center of the image, with the location drawn from a multivariate Gaussian with zero mean and diagonal covariance matrix $\sigma_{xx}^2=\sigma_{yy}^2 = 0.001$.
The magnitude of the source is chosen uniformly between [17-25].
The sources have independent Sersic indices drawn from a uniform distribution
\begin{equation}
    n_{\rm{ser}} \in U[0.5, 1]~,
\end{equation}
which controls the degree of curvature of the profile.
This yields a wide range of signal-to-noise ratios over which to train the network.

We note that a single Sersic profile light source is too simple to accurately model observed lenses.
We use this as a proof of principal (as done in other exploratory machine-learning based methods of subhalo detection).
Building machine-learning systems that are robust to source complexity is an area of active research.

{\textbf{Noise and detector effects}:}
As mentioned above, we use the HST module of the \texttt{SimulationAPI} in \texttt{Lenstronomy}.
We select the WFC3\_F160W camera band and use the built in pixel-based PSF.
The exposure time is set to 5,400~s for one orbit and 270,000~s for 50 orbits.
The sky brightness is set to magnitude 22.3 and the zero-point magnitude is 25.96.

With the input images and the target labels, we are able to train the network.
The specific model setup and the training details are described in more detail in the next section.

%%%%%%%%%%%%%%%%%%%%%%%%%%%%%%%%%%%%%%%%%%%%%%%%%%%%%%%%%%%%%%%%%%%%%% 
\section{Model architecture and training}
\label{sec:Model}
%%%%%%%%%%%%%%%%%%%%%%%%%%%%%%%%%%%%%%%%%%%%%%%%%%%%%%%%%%%%%%%%%%%%%% 

The technique of image segmentation is a type of object-detection method which aims at classifying every pixel in an image.
While there are many models for segmenting images, the best-performing ones have some similar features.
First, they are fully convolutional (in that there are no fully connected layers).
The best models also tend to have an encoder-decoder structure.
This allows the models to extract features across different scales and return a high-resolution segmentation map.
In particular, our network is based on the U-Net architecture~\citep{2015arXiv150504597R}, which has  excellent sensitivity to small objects in images.
For instance, the U-Net was first used to track cells in biomedical images.

In this work, we use a U-Net to classify each pixel in a strongly lensed image into one of 11 classes.
The classes are broken down as: belonging to the main lens, a subhalo with mass $\{10^{6},$ $10^{6.5}$, $10^{7}$, $10^{7.5}$, $10^{8}$, $10^{8.5}$, $10^{9}$, $10^{9.5}$, $10^{10}\}\msun$, or none of the above (which we will refer to as background throughout the paper).
While the goal of the network is classification, it allows us to both locate subhalos and obtain their mass.

Before an image is put through the network, the image is preprocessed by dividing by the maximum pixel value.
This normalization helps by forcing the brightness in all the images to have similar ranges.
Consequently, the network cannot base its classification on the absolute brightness of an image. 

\begin{figure}[t]
    \centering
    \includegraphics[width=\linewidth]{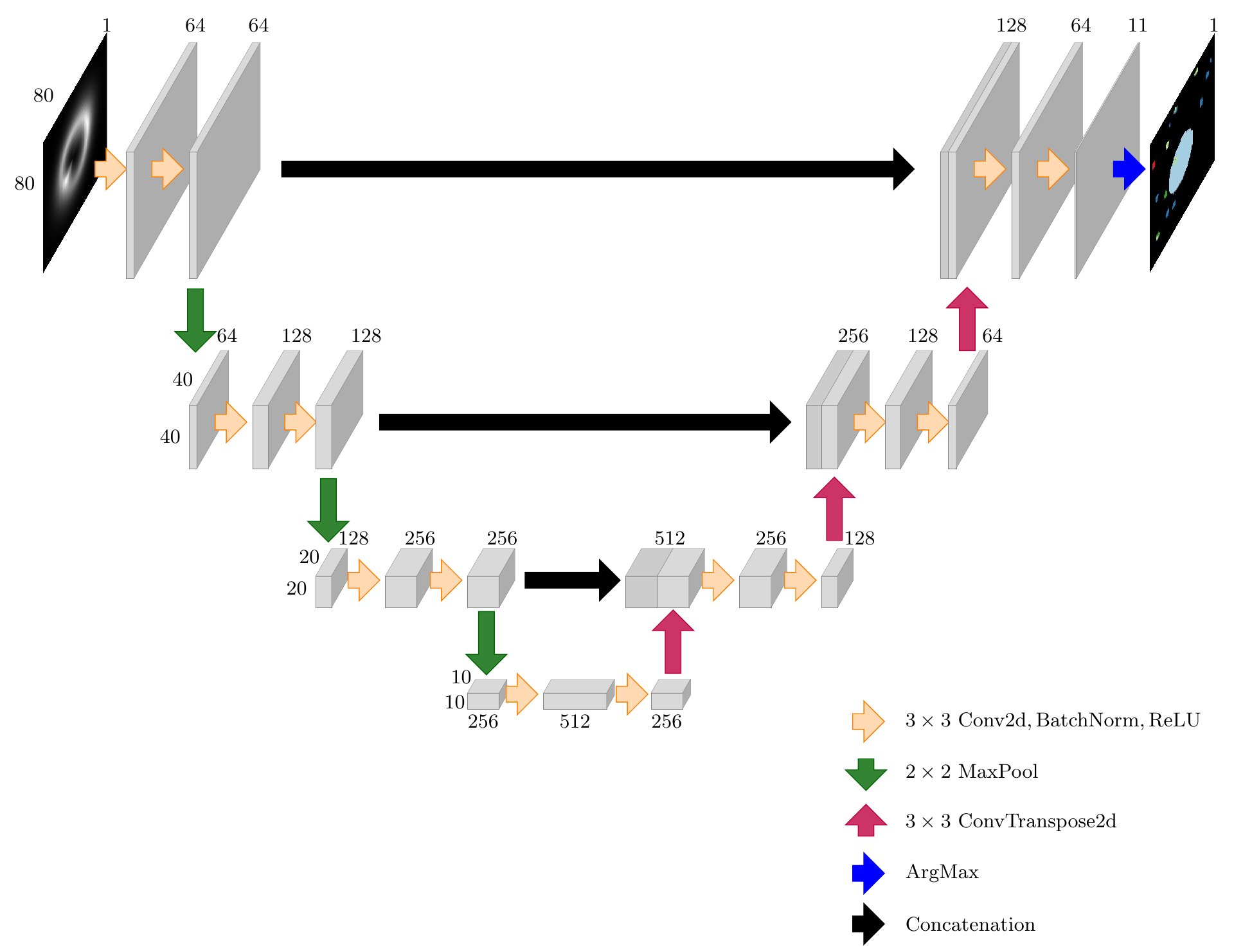}
    \caption{
    Network architecture. It takes in an $80\times80$ pixels image with a single layer and returns an image of the same size. 
    The pixel values in the output correspond to the predicted class. 
    }
    \label{fig:arc}
\end{figure}

Once the image has been normalized, it is ready to be segmented.
Our U-Net model architecture is implemented in \textsc{PyTorch}~\citep{NEURIPS2019_9015}, and Fig.~\ref{fig:arc} depicts our specific setup.
Each orange arrow represents three operations.
The first one is a 2D convolution in which a number of $3\times3$ pixels filters with learnable weights are convolved with the image.
The number of filters for each convolutional layer is denoted above each layer in the figure.
The second operation is batch normalization~\citep{2015arXiv150203167I}, which normalizes the data after the convolution, leads to faster training, and helps regularize the network.
The final operation represented by the orange arrows is applying the rectified linear unit (ReLU) activation function to the normalized data.
This is given by
\begin{equation}
    {\rm{ReLU}}(x) = \left\{\begin{matrix} 0, & x < 0 \\ x,& x>=0
\end{matrix}\right. ~.
\end{equation}
The convolutions in a given block are padded to preserve the number of pixels.

The green and red  arrows depict the down- and upsampling procedures,  which cut the number of pixels in half and double the pixel count, respectively.
The downsampling is done with a $2\times2$ maximum pooling operation.
The upsampling is done with a transposed convolution operation.
Note that the height and width of the data at each stage is marked at the beginning of each row in the figure.
Repeating the convolutional blocks (orange arrows) on the down sampled data with the same filter size allows the network to detect features at larger scales.
The upsampling transmits the information from these other scales back to the previous scale.
After the upsampling, the layer is concatenated with the last layer of the same height and width before downsampling (shown by the black arrows).
This allows the network to localize the new features and to avoid losing pattern information.

After the last convolution, our images have a depth of 11 channels corresponding to the 11 classes.
We apply the Softmax function along the channel such that the sum of a given pixel across all 11 channels is unity, and therefore its value for each channel can be thought of as a probability of belonging to the corresponding class.
Explicitly, this is given by
\begin{equation}
   \operatorname{Softmax} \left(z_i\right) = \frac{e^{z_i}}{\sum_{k=1}^{K} e^{z_k}}\equiv \widehat{p}_i  ,
    \label{eqn:softmax}
\end{equation}
where $z$ is the output for a pixel, the subscript denotes the pixel channel, and $K$ is the total number of channels: 11 for the problem at hand.
In this way, we interpret the channel to represent the predicted probability of belonging to a given class, denoted by $\widehat{p}_i$.

We train the network using a set of $9 \times 10^6$ images.
Of these images, $9 \times 10^5$ have only the source light and a smooth lens.
The remaining training images additionally contain exactly one subhalo. 
There are $9\times10^5$ images for each of the nine mass bins.
We use an independent set of  $5\times10^5$ images, with $5 \times 10^4$ from each of the sets mentioned above, to validate the model. 

As a classification problem, the cross-entropy loss per pixel is used.
This is given by
\begin{equation}
    L = \frac{-1}{n \times p} \sum_{i=1}^{n} \sum_{j=1}^{p} \sum_{k=1}^{K} y_{k}^{\left(i, j\right)}~\log\left(\widehat{p}^{\left(i, j \right)}_k \right),
\end{equation}
where the sum over $i$ goes over the $n$ images, the sum over $j$ runs over all of the $p$ pixels in an image, the sum over $k$ is the different possible classes, $y_k^{(i, j)}$ represents the true probability of pixel $j$ in image $i$ to belong to class $k$.
As the true pixel is either in a given class or not, $y_k^{(i, j)}$ is either 0 or 1.
Finally, $\widehat{p}$ is the probability predicted by the model.

We minimize the loss using the Adam optimizer~\citep{2014arXiv1412.6980K} with a learning rate of $10^{-3}$ and the default $\beta$ values.
The batch size is set to 100 images.
When the loss evaluated on the validation set has not improved for 5 epochs, the learning rate is dropped by a factor of 10, with a minimum rate of $10^{-6}$.
The training procedure is stopped when the validation loss has not improved for 15 epochs.

\begin{figure}[t]
    \centering
    \includegraphics[width=\linewidth]{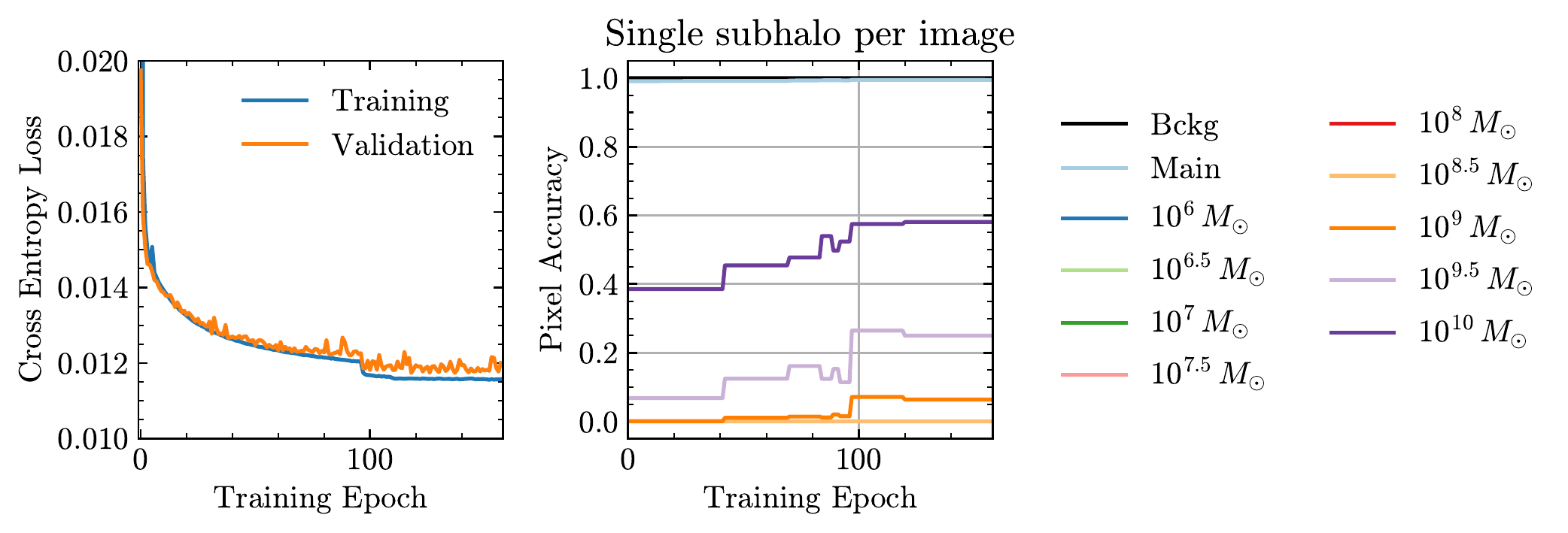}
    \caption{
    Example of training.
    The left panel displays the categorical cross-entropy loss as a function of the training epoch.
    The lowest validation loss occurs at after epoch 118, but the network is allowed to continue training until the validation loss has not improved for 15 epochs. 
    The learned parameters from the epoch with the lowest validation loss are used when applying the network to new data.
    The right panel shows the per-pixel accuracy of the validation data for each of the 11 classes for source magnitudes between 17 and 21.
    The background and smooth lens pixels are predicted correctly nearly $100\%$ of the time.
    The subhalos pixels are located and assigned the correct mass with varying accuracy for the different masses.
    Subhalos with $m\leq10^{8}\msun$, are never detected and the lines are overlapping at an accuracy of 0.
    We note that the accuracy of detecting subhalos is higher than the per-pixel accuracy (see Tab.~\ref{tab:NoThresholdPredictions}).
    }
    \label{fig:Training}
\end{figure}

An example of the training is shown in Fig.~\ref{fig:Training}.
The left panel shows the cross-entropy loss as a function of the training epoch, where the blue and orange lines denote the training and validation sets, respectively.
In addition to tracking the loss during training, we also compute the per-pixel accuracy of the validation data.
We define the per-pixel accuracy as
\begin{equation}
    \text{Pixel accuracy for class $k$} = \frac{\text{Number of pixels correctly predicted as class $k$}}{\text{Total number of truth level class $k$ pixels}}~.
    \label{eq:accuracy}
\end{equation}
For this, we define the a \emph{correct} pixel assignment when the class with the largest probability $\Big(\underset{k}{\rm{max}}~ \widehat{p}_k\Big)$ for a given pixel matches the true assignment.
There are 11 possible class assignments, so if the network is unsure of a given pixel's identity, all of the predicted probabilities could be around $1/11\sim 9\%$.
We emphasize that the pixel accuracy for the subhalo classes requires getting both the location (the pixel) and the mass of the subhalo correct.

The pixel accuracy for each class is shown as a function of the training epoch for the validation images with sources brighter than magnitude 21 in the right panel of Fig.~\ref{fig:Training}. 
In the first epochs, the model quickly learns to distinguish the main smooth lens from the background.
After this, the effects of the subhalos are recognized, starting with the heaviest, which have the largest effects on the image.
The accuracy for the $10^{10}\msun$ subhalo pixels reaches nearly $60\%$ by the end of training,.
The subhalos with mass $M\leq10^{8}$ almost never get predicted and thus have accuracies close to zero and are overlapping on the plot.
We note that at this stage we are only tagging pixels as belong to a subhalo (or not), but have not discussed the detection of a  subhalo as a whole. The eventual goal, however, is to build a catalog of subhalos with their positions and masses. This is done later in Sec.~\ref{sec:characterizing}.

With the training procedure defined, we now move on to examine the results of the network in more detail.

%%%%%%%%%%%%%%%%%%%%%%%%%%%%%%%%%%%%%%%%%%%%%%%%%%%%%%%%%%%%%%%%%%%%%% 
\section{Characterizing the network performance}
\label{sec:characterizing}
%%%%%%%%%%%%%%%%%%%%%%%%%%%%%%%%%%%%%%%%%%%%%%%%%%%%%%%%%%%%%%%%%%%%%% 
After the network has been trained, we apply it to a series of images that the network has not seen during training or validation to evaluate its out-of-sample performance.
We do this both for a network trained on images with a single HST-orbit-like noise and also for 50 orbits to investigate what could be gained with longer exposures.
First, we show an example output of the network, which helps to visualize the different channel probabilities and see common ways for the network to mislabel pixels.
We then compare the true target pixels to the predicted pixels to quantify the amount of correct and mislabeled pixels.
Finally, we run the network on images without substructure to determine the rate at which the network will claim to detect subhalos when they are not there (false positives).

%%%%%%%%%%%%%%%%%%%%%%%%%%%%%%%%%%%%%%%%%%%%%%%%%%%%%%%%%%%%%%%%%%%%%% 
\subsection{Example output}
\label{sec:example}
%%%%%%%%%%%%%%%%%%%%%%%%%%%%%%%%%%%%%%%%%%%%%%%%%%%%%%%%%%%%%%%%%%%%%% 

\begin{figure}[t]
    \centering
    \includegraphics[width=\linewidth]{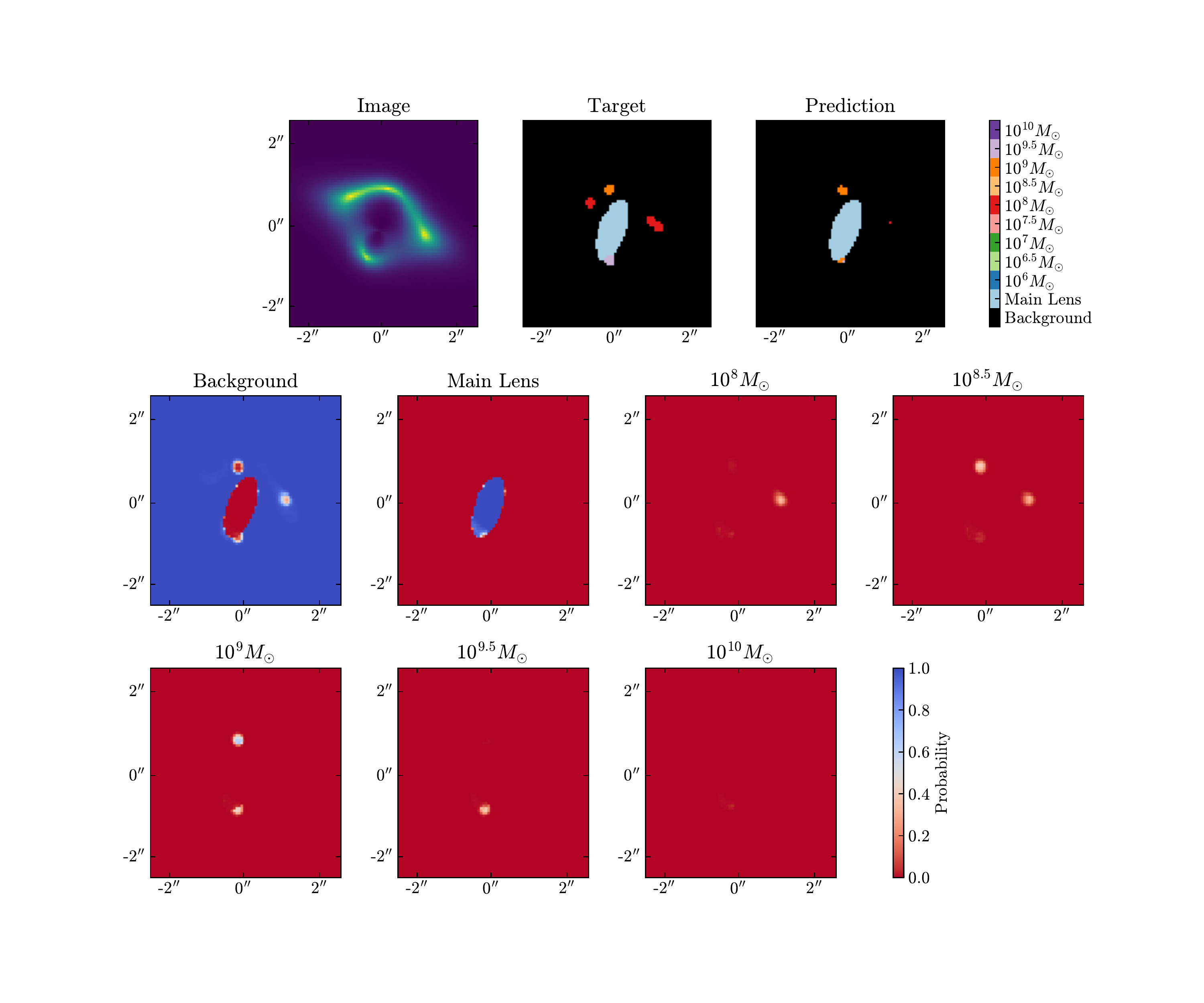}
    \caption{
    The top row shows (from left to right) an image with a simulated one orbit of HST exposure, which is input to the network, the target labels for each pixel, and the prediction from the network.
    The lower two rows show the probability assigned to each class for each pixel.
    The network prediction for each pixel is determined by the class with the maximum probability for that pixel.
    }
    \label{fig:example_detailed}
\end{figure}

In Fig.~\ref{fig:example_detailed}, we show a detailed example of the network applied to an image with multiple subhalos.
The upper row displays the observed image, which serves as input to the network (left), the truth level target labels (middle), and prediction (right).
We have assigned each pixel in the prediction image according to whichever class had the largest probability for that pixel.
The lower two rows display the individual class probabilities, with red representing low probabilities (the network is certain that the pixel does not belong to that class) and blue representing high probabilities (the network thinks this pixel belongs to the class).
We chose a color map such that probabilities near 50\% are white, showing that the network is unsure of those pixels.

First, we examine the network prediction for the main lens.
The shape and size of the lens is very similar between the target image and the predicted image.
The network picked up on the deviations from ellipticity of the lens. 
However, there are a few instances along the edge where the pixel assignments are incorrect (in the upper-left part of the main lens).
In the discussion below, these will be pixels that should be classified as background (main lens) but are misclassified as main lens (background).
Looking at the probability maps in the second row, for the background and main lens, we can see that the regions around the edge have class probabilities near 50\%.
Most of the pixel-wise errors made by the network come from edge effects like this.

In examining the target and prediction images of the top row, we see that the network successfully identifies three of the five subhalos in the image, despite their low masses and the overlap with the main lens.
The mass of the subhalo is indicated by the pixel colors.
The $10^9\msun$ subhalo is predicted to be the right size, but not all of the same pixels are correctly identified.
When examining the per-pixel accuracy in the next subsection, these will show up as pixels that should be predicted to be part of the subhalo (background), but are predicted to be in the background (subhalo) class.
Although some of the pixels were misidentified, the subhalo was still found and the pixel errors do not represent the network introducing spurious substructure.
If we examine the probability map for the $10^{9}\msun$ class (third row, first column), we see that the pixels around the edge of the upper subhalo are white, indicating that the network was not confident in these assignments.
In general, we find that sometimes the subhalos are predicted to be a few pixels too large and sometimes a few pixels too small, but just as in the preceding discussion, if a single pixel that should be assigned to a subhalo class is misclassified into the main lens or background, this does not indicate that the subhalo was not found.

This example contains another interesting feature; the two subhalos on the right side of the image are nearly overlapping.
The network is fairly certain that the pixels around them are not background.
However, it is uncertain whether these should be in the $10^8\msun$ or $10^{8.5}\msun$ bins.
Because we take the largest probability to be the prediction, and the subhalo mass predictions are split, all but one of the pixels are marked as background.
In principle, one could apply different thresholds, or use the probability map, depending on the particular use case.
Similarly, we see that the $10^{9.5}\msun$ subhalo at the bottom of the main lens has its probability spread out over many mass bins.
Because of this, the predictions cover two mass bins.
Despite this complication, it is clear that the network detected the subhalo.

%%%%%%%%%%%%%%%%%%%%%%%%%%%%%%%%%%%%%%%%%%%%%%%%%%%%%%%%%%%%%%%%%%%%%% 
\subsection{Testing on a single subhalo}
%%%%%%%%%%%%%%%%%%%%%%%%%%%%%%%%%%%%%%%%%%%%%%%%%%%%%%%%%%%%%%%%%%%%%% 

Now that the output of the network is better understood, we move on to quantify the the network's predictions.
The purpose of this is to determine what the network is predicting for the pixels of any class: are most of them correct? And if they are predicted wrongly, what class are they assigned to?
To assess this, we generate a new set of $5\times10^5$ images with the same amount of images with and without substructure as in the validation set: with $5\times10^4$ with no substructure and $5\times10^4$ for each mass bin with a single subhalo in the image.

\begin{figure}[t]
    \centering
    \includegraphics[width=0.7\linewidth]{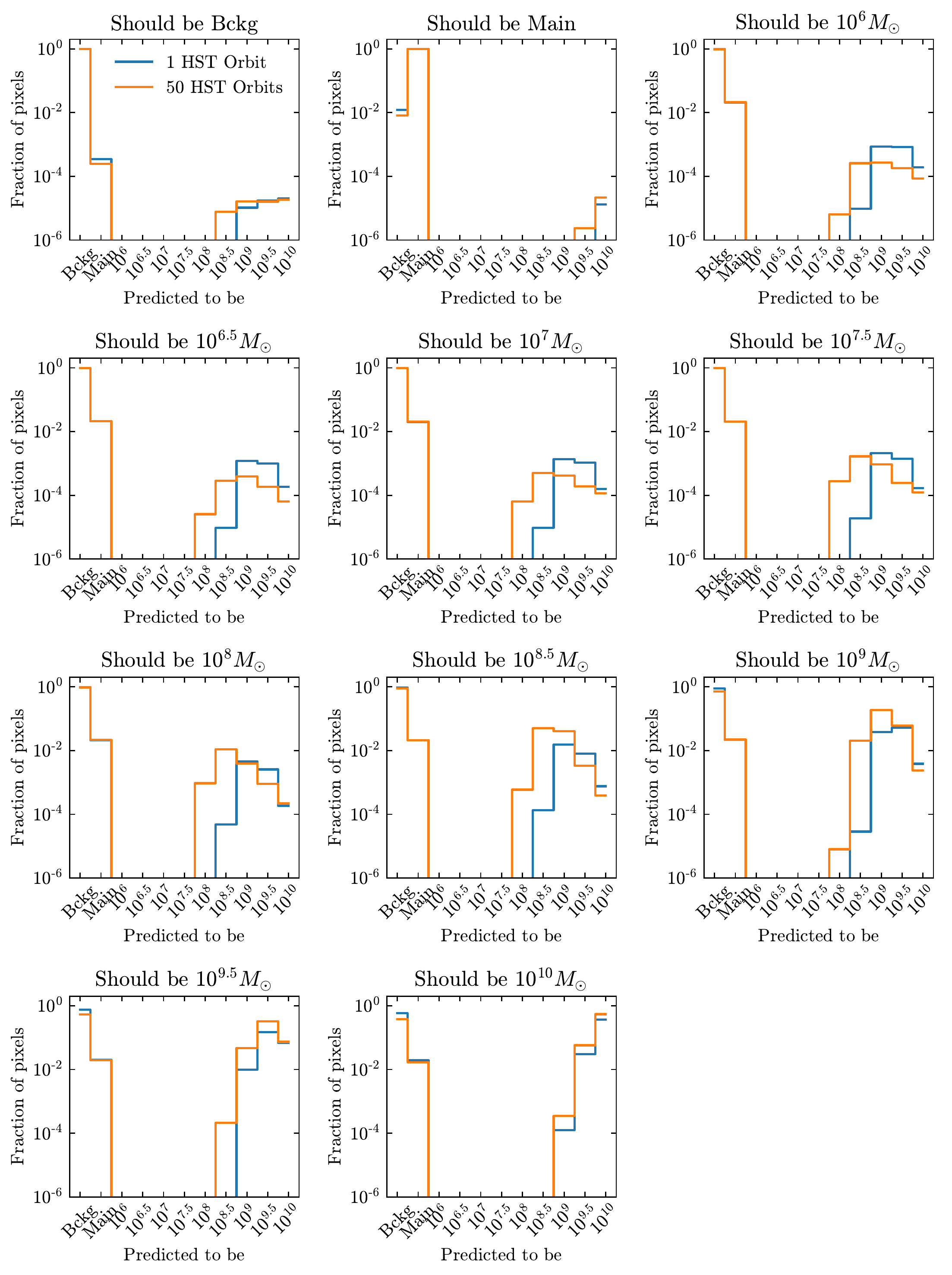}
    \caption{
    Each panel corresponds to pixels for which truth level are of the indicated class.
    The $x$-axis denotes the class that the pixels are predicted to be.
    The blue lines denote the network predictions on images with noise from 50 orbits of HST exposure and the orange is for images with a single orbit.
    Each line in each panel is normalized to unity.
    }
    \label{fig:PredVTrue}
\end{figure}

In Fig.~\ref{fig:PredVTrue}, we examine the per-pixel predictions on the images in the test set.
The title in each panel states the true label of the pixels, while
the $x$-axis denotes the predicted class.
The orange line corresponds to images with less noise (50 orbits) while the blue line corresponds to images with a single orbit of exposure.
Each panel is normalized such that the sum of all the classes is unity.

The first two panels (at the top, from left to right) show the pixels that at truth level correspond to the background or main lens classes, respectively.
The network often misclassifies a few pixels around the edge of the main lens (e.g. Fig.~\ref{fig:example_detailed} in the previous subsection).
This can either be the main lens being reconstructed as slightly too large or too small, or getting the shape slightly off, although it is typically only a handful of pixels.
Because there are so many background pixels, this corresponds to 0.01\% of the background pixels getting misclassified as the main lens over the entire test set; this is the most common type of misclassification for the background pixels.
Similarly, these errors around the edges of the main lens lead to $<1\%$ of the pixels that should be predicted as the main lens getting misclassified as the background class.

The rest of the class assignments in these two panels are roughly uniform for the classes which the network can detect.
The explanation is similar to that of the main lens itself: when the network locates a subhalo, some of the pixels around the edge can get mislabeled.
It is rare for the network to get the exact shape of the subhalo correct.
This was also shown in Fig.~\ref{fig:example_detailed}.
We emphasize that most of the pixels that are supposed to be background or the main lens, but are predicted to be a subhalo, do not represent additional false subhalos but are rather edge effects of this type.
More evidence of this is given in the next subsection.
Recall that the network does not  predict many pixels to be from subhalos with $m<10^{8.5}(10^8)\msun$ for a single (50) orbit(s).
This means that there are no edge effects associated to these classes (and consequently no pixels getting misclassified as a subhalo with $m<10^{8.5}(10^8)\msun$), as can be seen by the cutoff in the blue (orange) histograms. 

The edge effects can also be seen in the other panels.
For example, in the bottom-right panel, most of the pixels are correctly identified as belonging to a very massive subhalo.
However, some are incorrectly marked as the main lens or background.
These are from pixels around the edge of the subhalo.
In addition to edge effects, the network sometimes gets the mass wrong by one mass bin.
There is little difference in the results for the different exposures for the heavy subhalos.

Progressing toward the panels with the lighter subhalos, we notice that the fraction of pixels incorrectly labeled as background and/or main lens increases.
This makes sense as the magnitude of the deflection angles decreases with decreasing halo mass, so their effects are easier for the network to miss.

Fig.~\ref{fig:TrueVPred} in the Appendix shows the distribution of true labels for pixels predicted to belong to a given class, from which we see that in fact the predicted class is very likely to be correct.
Thus, if the network predicts a group of pixels to have the same class, it is very likely that a subhalo is present there.
The most common type of error for a pixel predicted to belong to a subhalo is that it should belong to a different (adjacent) mass bin.

Up to this point, we have only been discussing the per-pixel predictions.
This makes sense from a machine-learning perspective, but it does not necessarily address the physics goal of detecting subhalos themselves.
This brings to light one potential challenge of using image segmentation to detect substructure in images of strong lensing: how does one go from pixels to subhalos?
We have found that it is possible to either add extra pixels or miss pixels from a subhalo, especially around the edge.
However the per-pixel accuracies and the example shown in Fig.~\ref{fig:example_detailed} suggest that, on average, the size should be correct.
This means that we can get a subhalo count by summing the number of pixels predicted in each subhalo class and dividing by $4\pi ~\rm{pixels}/\rm{subhalo}$ (because we defined the target pixels such that the subhalos are comprised of a circle with a radius of 2 pixels).

\renewcommand{\arraystretch}{1.7}
\setlength{\tabcolsep}{5 pt}
\begin{table}[t]
    \centering
    \begin{tabular}{l | r r r r r r} 
        True Class & Not Detected &  $10^{8}\msun$
        & $10^{8.5}\msun$ & $10^{9}\msun$ & $10^{9.5}\msun$ & $10^{10}\msun$ \\
        \hline
        $10^{8}\msun$ & 99.2 [97.9]& 0.0 [0.1]& 0.0 [1.9]& 0.5 [0.2]& 0.3 [0.0]& 0.0 [0.0]\\
        $10^{8.5}\msun$ & 96.0 [78.1]& 0.0 [0.0]& 0.0 [14.7]& 2.7 [7.0]& 1.1 [0.2]& 0.2 [0.0]\\
        $10^{9}\msun$ & 83.2 [41.5]& 0.0 [0.0]& 0.0 [4.8]& 6.6 [44.6]& 9.8 [8.9]& 0.5 [0.2]\\
        $10^{9.5}\msun$ & 55.4 [14.7]& 0.0 [0.0]& 0.0 [0.0]& 1.2 [10.8]& 31.6 [65.5]& 11.8 [9.0]\\
        $10^{10}\msun$ & 27.1 [4.0]& 0.0 [0.0]& 0.0 [0.0]& 0.0 [0.0]& 5.7 [9.3]& 67.2 [86.7]\\
    \end{tabular}
    \caption{The percentage of subhalos predicted to each class when uniformly sampled source apparent magnitudes between 17 and 21. 
    The results for the network on images with 1 orbit and 50 orbits are indicated by the number without and with brackets.
    Note that in this test set, the subhalos were placed in bright pixels; the accuracy quickly decreases if the subhalos are not located near the Einstein ring.
    }
    \label{tab:NoThresholdPredictions}
\end{table}

We can then examine the subhalo detection accuracy over the test set, for the images with magnitudes between 17 and 21.
To define this accuracy, we take the pixels that at truth level belong to a given subhalo class and count the number of these pixels assigned to each of the 11 possible classes.
We then label the subhalo as belonging to the class with the largest count.
Using this notion of a subhalo detection, Table~\ref{tab:NoThresholdPredictions} shows how the subhalos of each mass bin were reconstructed.
The results for the network on images with noise from 1 orbit are indicated by the number without brackets and we include the images with less noise from 50 orbits as a reference in brackets.
For the images with an exposure of one orbit, the network finds 44.6\% of the subhalos with a mass of $10^{9.5}\msun$, of which 71\% are in the correct mass bin.
The subhalo detection accuracy, as well as the probability of getting the mass correct, increase for heavier subhalo masses.

Image segmentation struggles to resolve individual subhalos lighter than $10^{9}\msun$ because the noise is larger than the effect of the subhalos.
However, if the noise could be reduced with longer exposures or future telescopes, the range could be extended.
For instance, with 50 orbits, 21.9\% of the subhalos in the $10^{8.5}\msun$ bin are detected.

\begin{figure}[t]
    \centering
    \includegraphics[width=\linewidth]{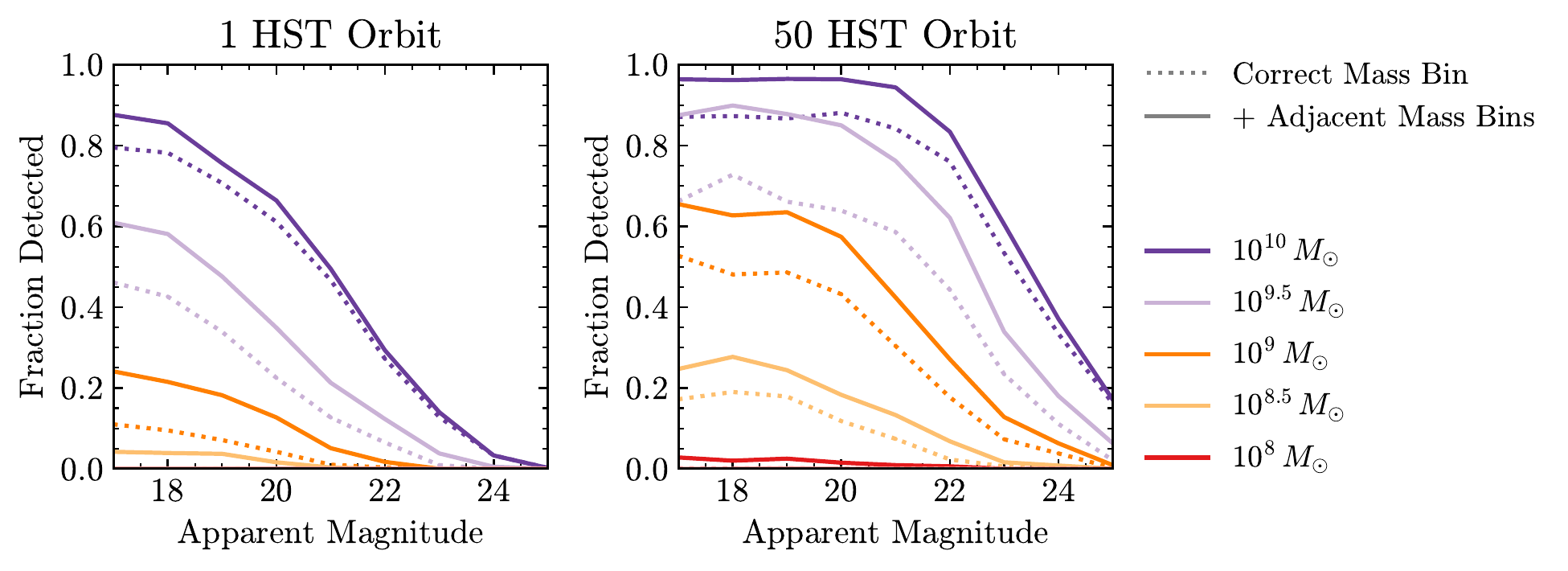}
    \caption{Fraction of subhalos detected as a function of the source magnitude.
    The dotted lines display the fraction of the subhalos that are predicted to belong to the true mass bin, and the solid line includes the neighboring mass bins.
    In this example, the source has a Sersic profile and the subhalo is placed in a bright pixel near the Einstein ring.
    For the dimmer sources, the signal-to-noise ratio is small and the subhalos are more challenging to detect.}
    \label{fig:MagnitudeDetection}
\end{figure}

These results indicate how important the signal-to-noise ratio in the images is for detecting substructure.
To assess this further, we simulate images following the same procedure, but specify the apparent magnitude of the source.
We generate $10^3$ images with a single subhalo for each fixed magnitude (17, 18, 19, 20, 21, 22, 23, 24, 25) for each subhalo mass bin.
The fraction of the subhalos that are detected for each mass bin as a function of the source magnitude is shown in Fig.~\ref{fig:MagnitudeDetection}.
The dotted lines display when the subhalo prediction matches the true mass bin, while the solid lines also include the adjacent mass bins.
Unsurprisingly, for very bright sources, the signal-to-noise ratio is high, and subhalos are easier to detect.
As the magnitude increases, the ratio of the noise increases.
The network cannot resolve subhalos that perturb the image below the level of the noise.

%%%%%%%%%%%%%%%%%%%%%%%%%%%%%%%%%%%%%%%%%%%%%%%%%%%%%%%%%%%%%%%%%%%%%% 
\subsection{Null tests}
%%%%%%%%%%%%%%%%%%%%%%%%%%%%%%%%%%%%%%%%%%%%%%%%%%%%%%%%%%%%%%%%%%%%%% 

In the last section, we claimed (in Fig.~\ref{fig:PredVTrue}) that the flatness of the distribution for the pixels that at truth level should belong to the background class, but instead get classified as subhalos, is evidence that this type of error is dominated by edge effects.
In part, this is because the test set that we are analyzing had the same number of subhalos in each mass bin.
Additionally, Fig.~\ref{fig:example_detailed} provided an example of one such case. 
In this section, we provide further evidence of this claim in a systematic way, by determining the rate at which the network finds spurious substructure in images where there is only a smooth lens.

\begin{figure}[t]
    \centering
    \includegraphics[width=\linewidth]{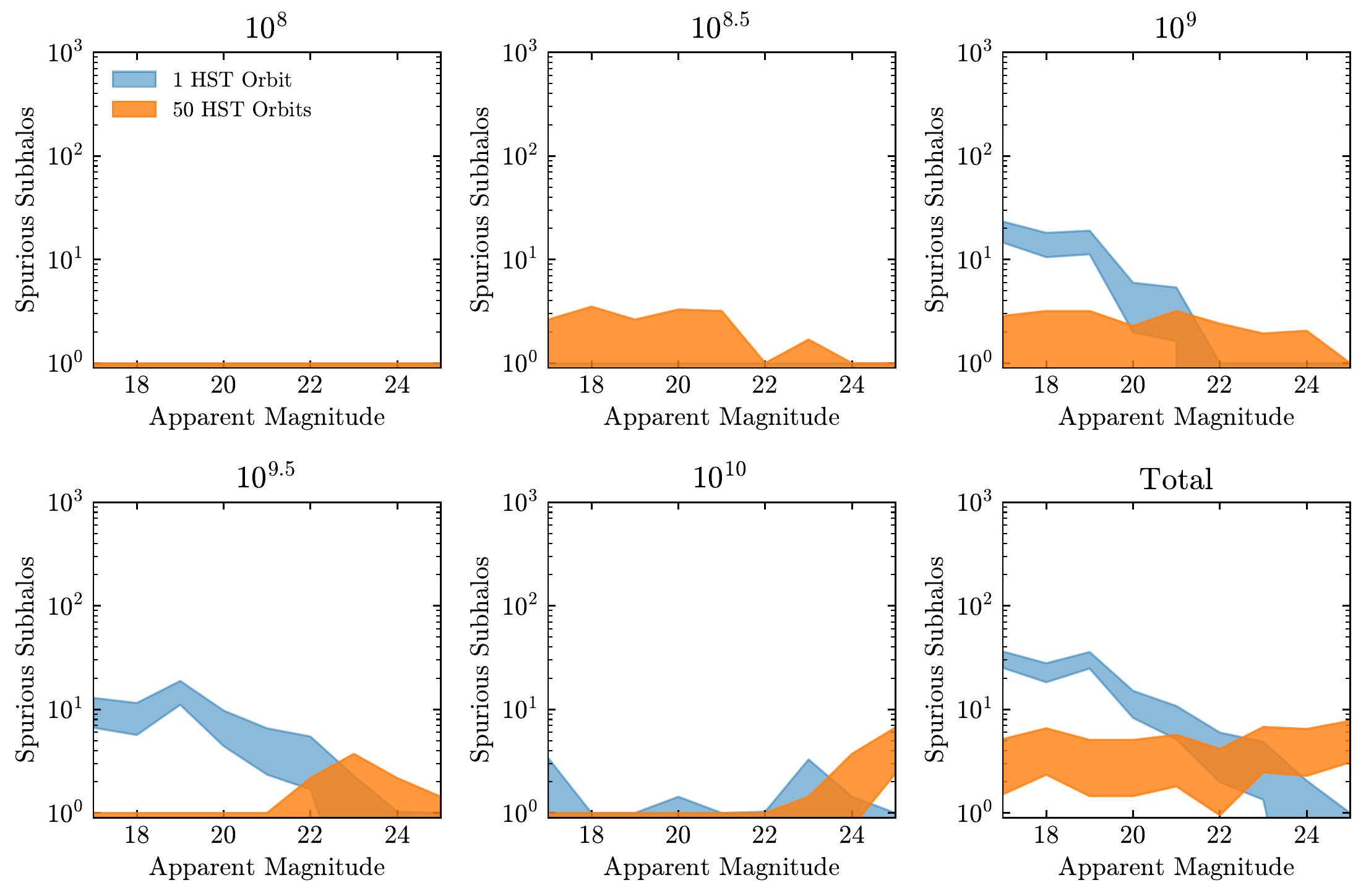}
    \caption{1000 images are generated for source magnitudes with integer values in the range of 17-25 with no substructure in the lens. 
    The panels show the number of times that the network predicted a subhalo to be in the image, with the uncertainty estimated by the square root of the counts.
    The total rate of spurious predictions by the network is less than 30 subhalos per 1000 images.
    }
    \label{fig:NullTest}
\end{figure}

We use the same dataset as in the magnitude test of the previous section, but we only examine the subset of images with no substructure.
As the signal-to-noise changes as a function of the apparent magnitude, so does the network's ability to identify substructure.
There are $10^3$ images for each magnitude value and the resulting counts are shown in Fig.~\ref{fig:NullTest}.
The error bars were estimated using the square root of the estimated subhalo count, $\sqrt{N_{\rm subhalo}^{\rm{predicted}}}$. 

 The network struggles to detect $10^8\msun$ subhalos, but it knows this and therefore does not predict any spurious ones in that mass bin either.
Similarly, for the $10^{8.5}\msun$ bin, the network trained on images with one orbit of exposure does not detect these subhalos and does not predict spurious ones.
However, with reduced noise of 50 orbits, the network starts to detect them, and thus sometimes has false predictions.

As the mass increases, it becomes easier for the network to detect the subhalos.
We note a similar interesting trend, specifically in the 1 orbit data:
the spurious subhalo rate decreases as the brightness goes down.
Initially this is counter-intuitive, as the signal-to-noise ratio goes up, the fake rate goes down.
However, the network knows that it cannot detect subhalos with the given signal-to-noise ratio, so it does not try to predict them.
In the images with 50 orbits of data, the noise is small enough that the different magnitudes are not significantly different. From this, we observe that the spurious subhalo rate does not change much.

The last panel (bottom right) shows the total sum of all of the subhalos.
Across a wide range of image brightness (and signal-to-noise ratio), there are fewer than 30 spurious subhalos per 1000 images.

%%%%%%%%%%%%%%%%%%%%%%%%%%%%%%%%%%%%%%%%%%%%%%%%%%%%%%%%%%%%%%%%%%%%%% 
\subsection{Testing on images with many subhalos}
\label{sec:manysubhalos}
%%%%%%%%%%%%%%%%%%%%%%%%%%%%%%%%%%%%%%%%%%%%%%%%%%%%%%%%%%%%%%%%%%%%%% 

\begin{figure}[t]
    \centering
    \includegraphics[width=\linewidth]{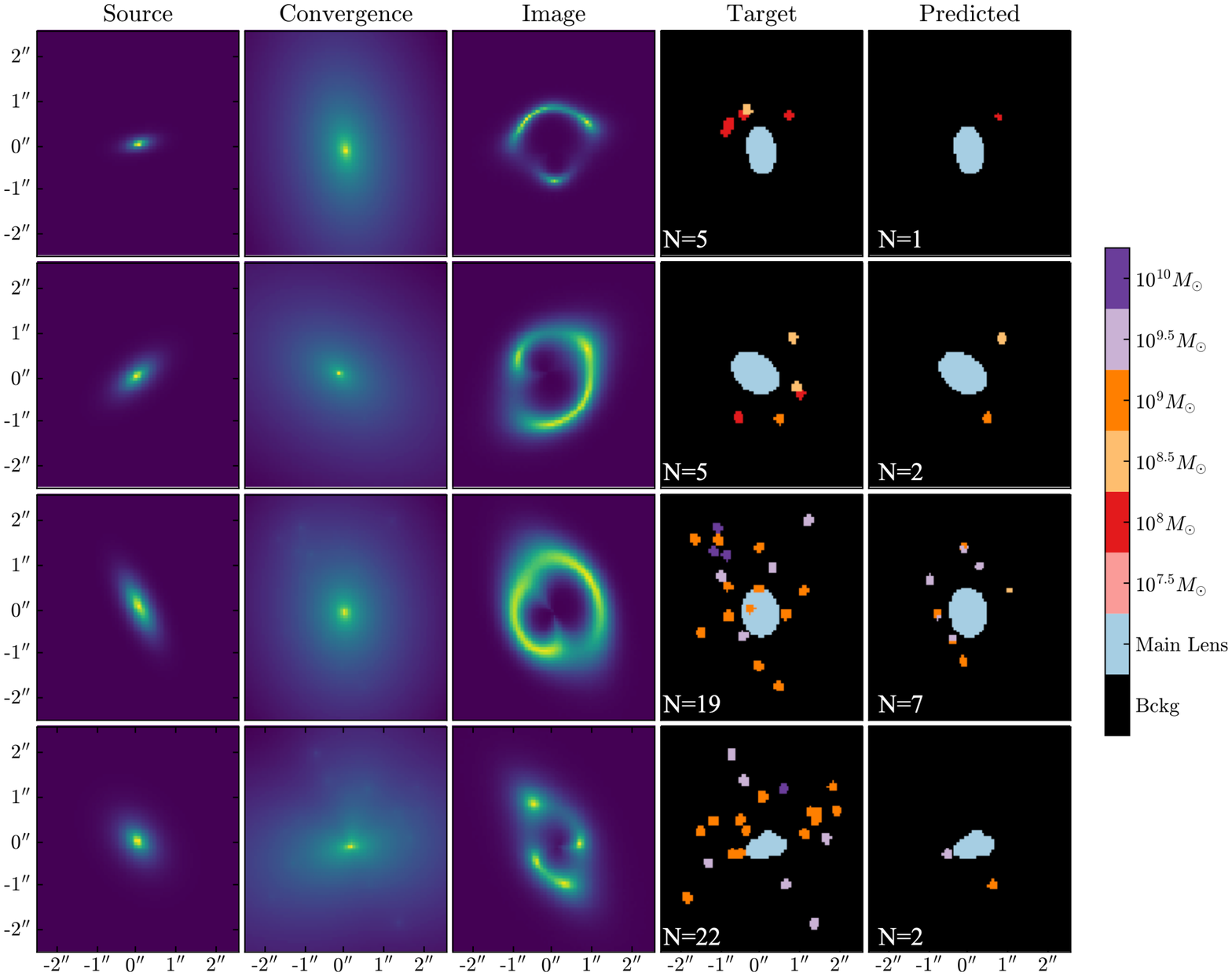}
    \caption{
    The network trained on images with a single subhalo is now applied to images with rich distributions of substructure.
    The network has low sensitivity to subhalos that are far from the bright pixels around the Einstein ring.
    Subhalos that are near each other can get reconstructed as a single subhalo with their combined mass.
    The number of true/predicted subhalos are denoted in the figure.
    }
    \label{fig:immanysubhalos}
\end{figure}

Up to this point, we have presented results from networks trained on images with at most a single subhalo near the brightest pixels.
However, galaxies are expected to have a large population of subhalos distributed throughout the main dark matter halo.
We therefore investigate whether, without knowing this true distribution of subhalos, our network can be used to identify a population of substructure (i.e. more than a single subhalo per image).

To test this, we generate a set of images with up to 25 subhalos (with the actual number in each image drawn from a uniform distribution) with random masses and locations.
Four of these images are shown in the panels of the third column of Fig.~\ref{fig:immanysubhalos}.
The fourth column shows the true pixel labels and the final column shows the network output.
As a reference, we include the source light flux and the log of the convergence in the first and second columns, respectively.
We denote the number of subhalos in each of the target and predicted images.
All of these images include noise from 50 orbits.

The network predictions for these images illustrate that substructure far from the images is hard for the network to capture.
Each row also illustrates different interesting properties.
In the top row, we see the network detect a light subhalo in the $10^{8}\msun$ bin.
The second row contains an example of the network detecting two subhalos, but missing some of the lighter subhalos in the image.
In the third and fourth rows, subhalos are inside of the Einstein radius and are missed by the network.
The interior of the ring is also far from the light, making detection harder.
Finally, in the last row, we see an example of two subhalos in the $10^{9}\msun$ mass bin being near each other, and the network detects them as a subhalo in the $10^{9.5}\msun$ bin.

From these examples, we see that the network trained on images with a single subhalo is able to detect a population of substructure.
As the subhalos get further away from the bright pixels, the network loses sensitivity to them.
The presence of subhalos close together is also a challenge for the network.
When there is a hierarchy between their masses, the heavier subhalo will wash out the effects of the smaller one.
This results in only a detection of the heavy subhalo.
If nearby subhalos are similar in mass, they may get reconstructed as a single subhalo in the bin corresponding to the sum of their masses. 

The variety of subhalo populations that were detected in Fig.~\ref{fig:immanysubhalos} (both in terms of raw number of subhalos as well as their mass distribution) also highlights one of the strengths of our single subhalo training methodology.
We found empirically that when we instead trained a network on images with many subhalos drawn from a power law mass function and then tested it on images with populations of subhalos drawn from a mass function with a different power law index, the detection accuracy became biased.
For instance, if the network is trained on images with many subhalos drawn from a steep mass function, the few images with heavy subhalos will almost certainly have lots of light substructure as well.
When we applied the network to images generated with a shallower mass function, the false-positive rate was higher for images with heavy subhalos: when it saw a heavy subhalo, it expected more light subhalos than there were, thus introducing false substructure.
By only having a single subhalo in the training images, our fiducial network never learns to make decisions based on population characteristics, and thus generalizes well to images with very different subhalo populations.

We have now shown the our network can detect subhalos near the Einstein, and is capable of generalizing to images with different subhalo populations.
Additionally, it has a very low false-positive subhalo rate of three subhalos per 100 images.
In the next section we demonstrate how such a network could be used to extract the SMF from an ensemble of strong lens images.

%%%%%%%%%%%%%%%%%%%%%%%%%%%%%%%%%%%%%%%%%%%%%%%%%%%%%%%%%%%%%%%%%%%%%% 
\section{Determination of the SMF}
\label{sec:shmf}
%%%%%%%%%%%%%%%%%%%%%%%%%%%%%%%%%%%%%%%%%%%%%%%%%%%%%%%%%%%%%%%%%%%%%% 

The population of subhalos under a CDM scenario is found to be well described by a power law of the form
\begin{equation}
\frac{dN}{dm} \propto m^{\beta},
\label{eqn:shmf}
\end{equation}
with $\beta=-1.9$~\cite{2008MNRAS.391.1685S}.
However, models beyond CDM can affect this SMF. In previous sections, we have described how our U-Net model is able to accurately detect subhalos in simulated images. In this section, we show that we can determine the SMF using the network outputs from many images.

We first generate a mock catalog of images that can have more than a single subhalo.
In each image, we draw the number of subhalos from a uniform distribution between 0 and 25, and place them uniformly throughout the image~\cite{2018PhRvD..97b3001D, 2018PhRvD..98j3517D}.
The masses of the subhalos are drawn from a power law with fixed $\beta=-1.9$.
As the network with noise is not able to detect substructure with $m < 10^8\msun$, we do not generate subhalos with masses that would end up in the $m\leq 10^{7.5}\msun$ bins.
Similarly, we only generate images with sources brighter than magnitude 21 since the network struggles to detect substructure in dimmer images.
Our goal will be to infer the value of $\beta$ from this catalog.

\begin{figure}[t]
    \centering
    \includegraphics[width=11cm]{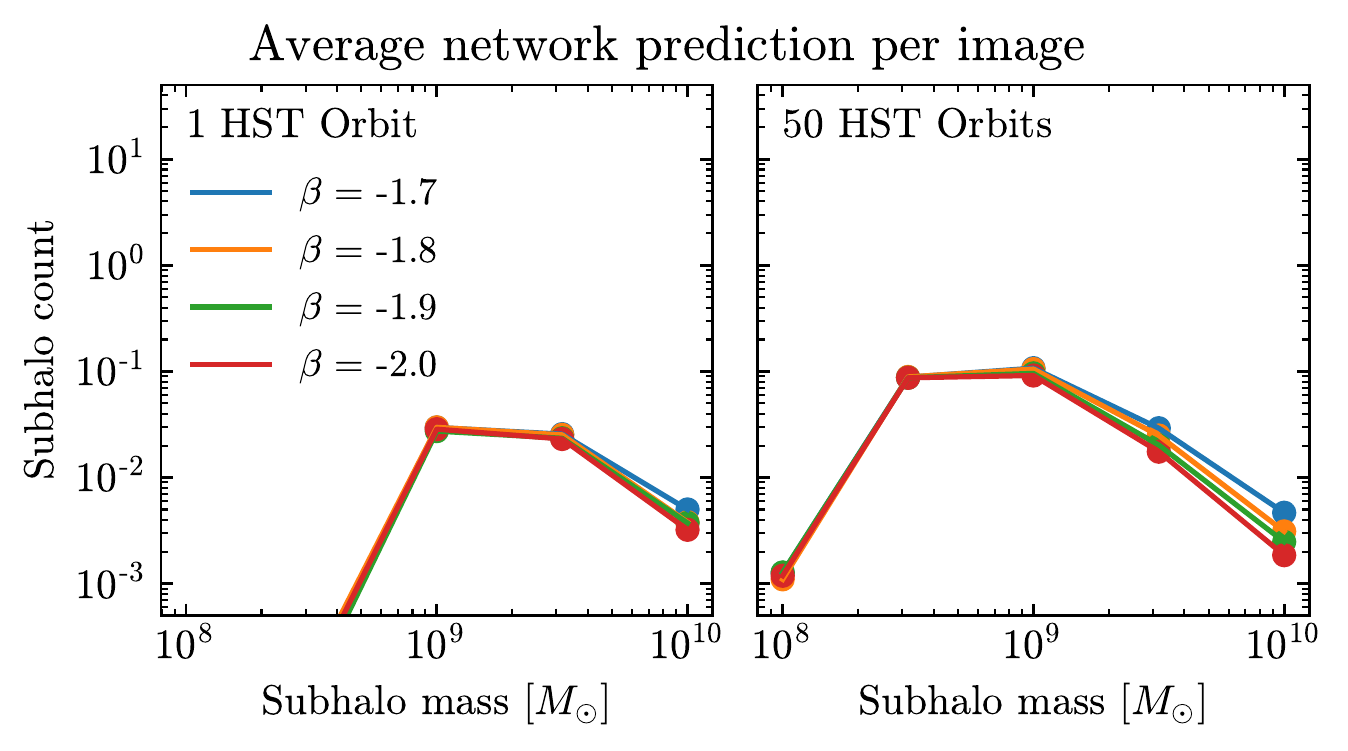}
    \caption{The left and right panels display the average number of detected subhalos per class for $10^5$ images with one and 50 orbits of HST exposure, respectively.
    The images can have between 0 and 25 truth level subhalos uniformly distributed throughout the image.
    The masses of the subhalos are drawn from a power law with an index denoted by $\beta$.
    The network has less sensitivity to lower mass subhalos, resulting in the curved shape.
    To infer the power law index of an independent dataset, we find which value of $\beta$ yields a curve closest to that observed in the data.
    }
    \label{fig:LikelihoodIntro}
\end{figure}

If we were able to perfectly reconstruct every subhalo, we could just fit the counts per bin to the functional form in Eq.~\eqref{eqn:shmf}.
However, the accuracy is worse for lower mass subhalos, effectively changing the shape of the extracted mass function.
An example of this is shown in Fig.~\ref{fig:LikelihoodIntro}.
Here we simulated $10^5$ images, using the same prescription as in our mock catalog, for $\beta=$-1.7, -1.8, -1.9, and -2.0.
We then applied the network to each image and counted how many pixels were assigned to each class.
This pixel count was converted to an approximate subhalo count by dividing by $4\pi$ (because the truth labels are defined by a circle with a radius of 2 pixels).
The left panel shows the average number of detected subhalos for the images with one orbit, and the right panel shows the average counts for the images with 50 orbits.
By placing the subhalos randomly (instead of only in bright pixels) along with the fact that most of the subhalos are light, we see that the average image does not have a subhalo get detected.
Here, we must emphasize that the detection performance presented above was for subhalos only in bright pixels near the Einstein ring.

Fitting the data with Eq.~\eqref{eqn:shmf} would not make sense because the reconstructed data does not lie along a fixed power law.
Instead, we infer the most probable power law index to have generated the observed subhalo counts extracted from a set of images with our U-Net model.
To do so, we build a likelihood function which is the product of Gaussian likelihoods for each mass bin.
Namely, the likelihood is given by
\begin{equation}
    L(\beta) \equiv \prod_{i\in \text{mass bins}} G\Big(o_i \Big| \mu_i(\beta), \sigma_i(\beta)  \Big),
    \label{eqn:likelihood_initial}
\end{equation}
where $o_i$ is the observed number of detected subhalos in bin $i$, $\mu_i(\beta)$ is the expected value, and $\sigma_i(\beta)$ is the estimated standard deviation given $\beta$.
To use the likelihood to infer $\beta$, we need to derive $\mu_i(\beta)$ and $\sigma_i(\beta)$.
This is done in the next subsection.
We emphasize that these derivations will include details that are specific to our mock catalog of images, and may not generalize to the real world, but we will discuss how to generalize to other catalogs.

% **********************************
\subsection{Expectation and variance}
% **********************************

Here, we derive the expectation and variance used in the likelihood function given in Eq.~\eqref{eqn:likelihood_initial}.
First, we define the efficiency to tag a subhalo as
\begin{equation}
    \varepsilon_i = \frac{\text{Number of subhalos predicted in class $i$}}{\text{Number of true subhalos in class $i$}}.
\end{equation}
Because we train our network on images with a single subhalo, the network does not know about population-level statistics, which is why the efficiency to zeroth order is not a function of $\beta$.
If an image has $N$ real subhalos, we define $\varphi$ as the true fraction of subhalos in each mass bin, $N_i$.
Note that $\varphi$ is necessarily a function of $\beta$ and is given by
\begin{equation}
    N_i(\beta) = \varphi_i(\beta) ~N,
\end{equation}
where $N_i$ is the true number of subhalos in the $i$th bin.
Thus, the number of subhalos expected to be predicted in a given class for an image with $N$ total true subhalos can be written as
\begin{equation}
    \langle n_i (\beta)| N\rangle = \varepsilon_i~\varphi_i(\beta)~N.
    \label{eqn:expect1}
\end{equation}
where $n_i$ is the number of predicted subhalos in bin $i$.

\begin{figure}[t]
    \centering
    \includegraphics[width=0.8\linewidth]{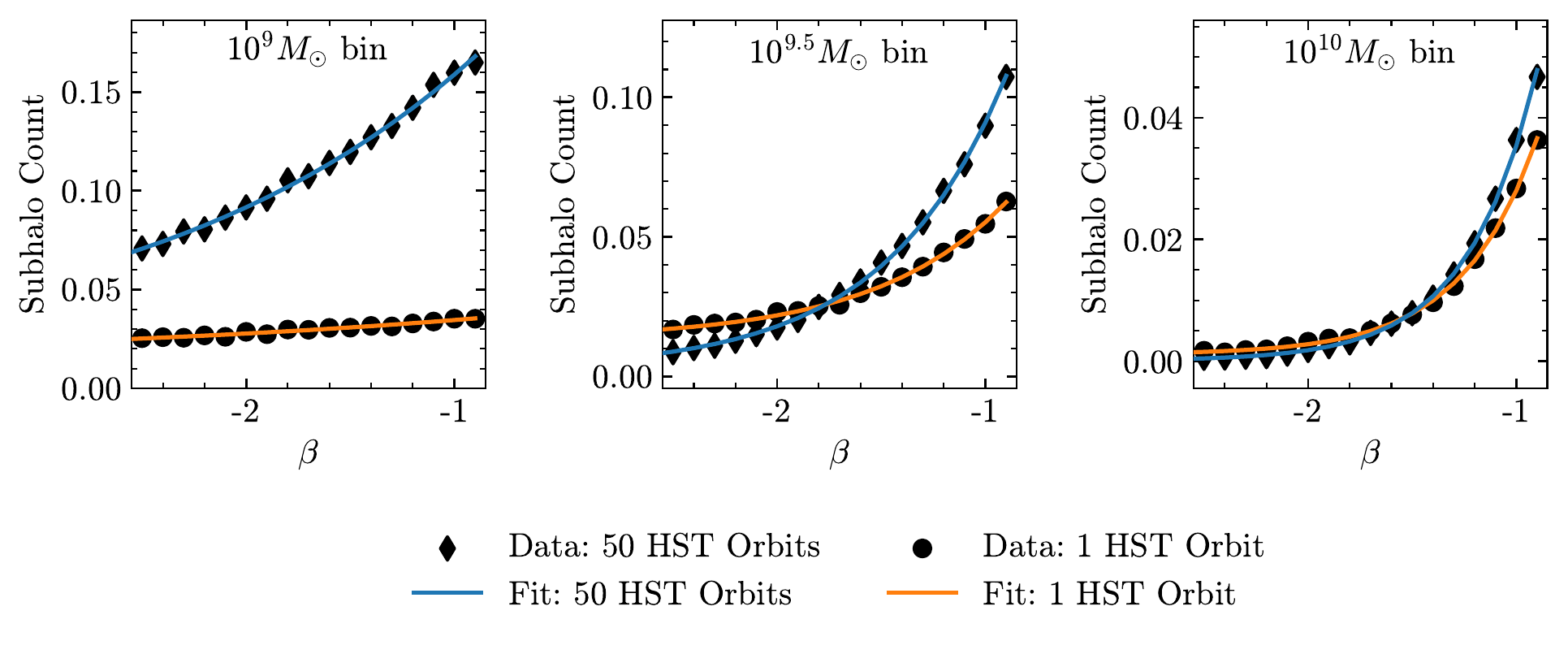}
    \caption{
    Data points show the average number of subhalos detected by our network per image over $10^5$ sample images with different exposure lengths.
    The lines show exponential fits to the data, as described in the text.
    These are used to define the expected subhalo counts and the standard deviations used in the likelihood function.}
    \label{fig:LikelihoodTemplates}
\end{figure}

By passing $10^5$ images through the trained network, we obtain a good estimate for
\begin{eqnarray}
    \epsilon_i(\beta) &=&  \varepsilon_i~\varphi_i(\beta)~ \nonumber \\
        &=&\langle n_i (\beta)| N\rangle / N ~,
\end{eqnarray}
where $\epsilon_i(\beta)$ is essentially the rate at which subhalos are detected in mass bin $i$ given the value of $\beta$ over the range of apparent magnitudes.
For example, in Fig.~\ref{fig:LikelihoodTemplates} we applied the network to count the number of predicted subhalos in $10^5$ images for various fixed values of $\beta$ with an average of 12.5 subhalos per image.
Rather than showing the results as a function of the subhalo mass bin (as done in Fig.~\ref{fig:LikelihoodIntro}), we show the average count per image as a function of $\beta$, with each mass bin shown in a different panel.
We find that the data can be fit well with an exponential function of the form
\begin{equation}
    \epsilon_i(\beta) = a_i + b_i\, e^{c_i \beta}~.
\end{equation}

Given the rate ($\epsilon_i$) and the true number of subhalos ($N$) in a single image, the number of predicted pixels in a given class should be be Poisson distributed about the expectation.
Thus, the probability ($p$) to predict $n_i$ subhalos is given by
\begin{equation}
    p\big(n_i (\beta)\big) = P\big(n_i | \epsilon_i(\beta)~N\big) = \frac{\big(\epsilon_i(\beta)~N \big)^{n_i} e^{-\epsilon_i(\beta)~N}}{n_i!}~.
\end{equation}
The expected number of detected subhalos in a single image with $N$ true subhalos from Eq.~\eqref{eqn:expect1} can then be rewritten in terms of the the weighted sum over the individual probabilities
\begin{equation}
    \langle n_i(\beta)| N\rangle = \sum_{n_i} n_i~p\big(n_i (\beta)\big)~.
\end{equation}

We now build in the assumptions about our mock data.
In each image, we have placed between 0 and $N_{\rm max}=25$ true subhalos, with the number drawn from a uniform distribution.
The expected number of detected subhalos in class $i$ for any image is then the average over the $N_{\rm max} + 1$ possible values (including 0 subhalos),
\begin{equation}
    \langle n_i(\beta) \rangle = \sum_{N=0}^{N_{\rm max}} \frac{1}{N_{\rm max} +1} \sum_{n_i} n_i~p\big(n_i (\beta)\big)~. 
    \label{eqn:expect2}
\end{equation}
In an analysis of real data, the actual distribution of subhalos would need to be obtained through detailed $N$-body simulations.
Similarly, the brightness of the sources in the simulated calibration set should resemble the real data to be analyzed.

In any given image, the number of subhalos in a bin can vary widely.
To compute the variance, we calculate $\langle n_i^2 \rangle$ as a function of $\beta$, given by
\begin{equation}
    \langle n_i^2 (\beta)\rangle = \sum_{N=0}^{N_{\rm max}} \frac{1}{N_{\rm max} + 1} \sum_{n_i=0}^{N} n_i^2~p\big(n_i (\beta)\big)~. %P\left(n_i | \epsilon_i(\beta)~ N\right)~.
\end{equation}
Then the variance is
\begin{equation}
    \sigma_{n_i}^2(\beta) = \langle n_i^2(\beta) \rangle - \langle n_i(\beta) \rangle^2~.
\end{equation}
The range of possible true subhalos makes this variance large.

Up to this point, we have examined the expectation and variance for the number of detected subhalos per bin in a single image.
However, to determine the SMF we analyze many images and sum the resulting subhalo detections for each mass bin..
The expected number of detected subhalos and the variance for $N_{\rm{images}}$ independent images is given by
\begin{eqnarray}
     \langle n_i (\beta) | N_{\rm{images}}\rangle &=& N_{\rm{images}} \langle n_i(\beta) \rangle;
     \label{eqn:expect_final}\\
     \sigma^2_{n_i(\beta) | N_{\rm{images}}} &=& N_{\rm{images}} \sigma^2_{n_i}(\beta).
     \label{eqn:variance_final}
\end{eqnarray}
The value of the mean and standard deviations for each bin included in the likelihood function are then given by
\begin{equation}
    \mu_i(\beta) = N_{\rm{images}} \langle n_i(\beta)\rangle
\end{equation}
and
\begin{equation}
    \sigma_i(\beta) = \sqrt{N_{\rm{images}} \sigma^2_{n_i}(\beta)}~.
\end{equation}

% **********************************
\subsection{Inferring the Power Law}
% **********************************

At this stage, given a SMF power law index and some number of images, we can compute the expected number of detected subhalos and the standard deviation for each bin.
When we apply our network to a set of images, we can then determine which value of $\beta$ yields expectations closest to the observed detections using the likelihood function defined in Eq.~\eqref{eqn:likelihood_initial}.
Recalling that the Gaussian expectation and variance are a function of $\beta$ and that the observed data does not change, the likelihood is then a function only of $\beta$,
\begin{equation}
    L = L(\beta).
\end{equation}
To test a hypothesized value of $\beta$, the likelihood ratio is used,
\begin{equation}
    \ell(\beta) = \frac{L\big(\beta\big)}
    {L\big(\widehat{\beta}\,\big)},
\end{equation}
where $\widehat{\beta}$ is the value of the slope which maximizes the likelihood.
To find $\widehat{\beta}$, we use the Nelder--Mead~\citep{Nelder:1965zz} algorithm implemented in \textsc{SciPy}~\citep{2020SciPy-NMeth}.
We then compute the test statistic $t_{\beta}$, defined as 
\begin{equation}
    t_\beta = -2 \log\ell\left(\beta\right),
    \label{eqn:tbeta}
\end{equation}
with which we can determine confidence intervals.

\begin{figure}[t]
    \centering
    \includegraphics[width=0.9\linewidth]{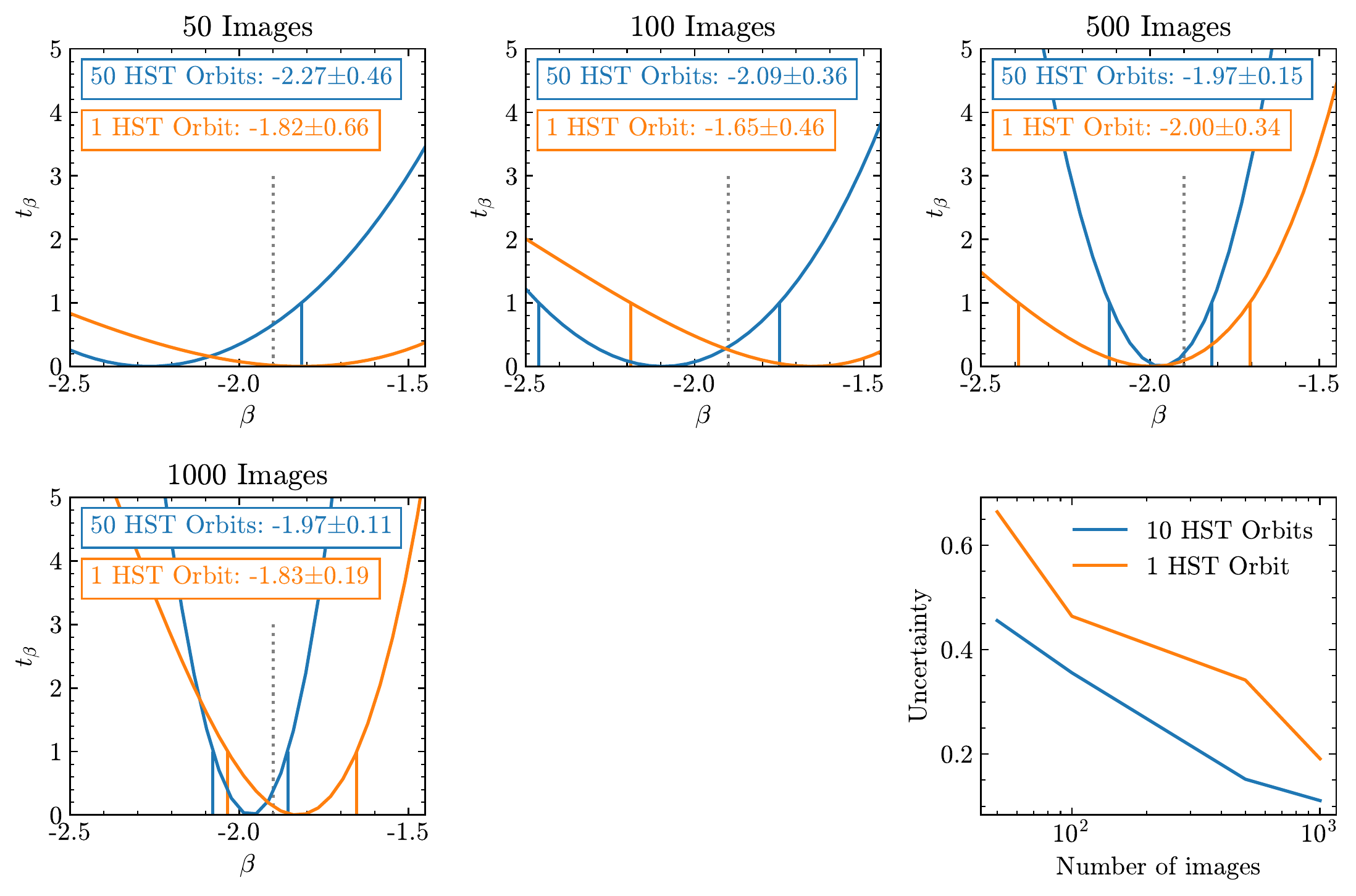}
    \caption{
    Result of scanning the test statistic $t_{\beta}$ when scanning over the power law index $\beta$.
    The test statistic is defined in Eq.~\eqref{eqn:tbeta}.
    The value of $\beta$ that minimizes $t_{\beta}$ produces expected count most similar to that in the mock observed data.
    The true value of $\beta$ in the mock data is -1.9 and is marked by the gray dotted line.
    The 1$\sigma$ confidence interval encompasses the values of $\beta$ between $t_{\beta}=1$.
    The last panel shows that the uncertainty decreases as the number of images increases.
    The orange and blue lines shown the results for image with one and 50 orbits of exposure, respectively.
    The last panel shows the 1-$\sigma$ uncertainty.
    The uncertainty on the best fit for the images with a single orbit is about 25-50\% larger than the when using images with 50 orbits.
    }
    \label{fig:Likelihoods}
\end{figure}

Fig.~\ref{fig:Likelihoods} does this for images containing subhalo populations drawn from an SMF with $\beta_{\rm{true}}=-1.9$, for different numbers of images in each panel.
The best fit is the value of $\beta$ that minimizes $t_{\beta}$, while the $1\sigma$ uncertainty is given by the range between $t_{\beta} = 1$.
The orange and blue lines denote the results from images with noise from one and 50 orbits of exposure, respectively.
The last panel shows the spread of the confidence intervals as a function of the number of images used for the fit.

The first panel uses only 50 images, and the resulting inferred values have large uncertainties.
This is primarily due to the large variance in the number of subhalos per image, which is incorporated into the uncertainty.
Increasing the number to 100 images, the uncertainty drops by a factor of 1/3.
The trend continues as more images are added, with the best fit converging to the true value and the uncertainty decreasing.

We again note that decreasing noise increases our ability to tag low-mass substructure.
The uncertainty on the inferred value of $\beta$ can be decreased by around $20\%$ (without including the possible detection of even lower mass subhalos).
Overall, we find that when measured over three mass bins from $10^9\msun$ to $10^{10} \msun$ the SMF slope is recovered with an error of 36 (20)\% for 50 images, and this improves to 10 (5.5)\% for 1000 HST-quality images with one (50) orbits of exposure.

The results presented here strongly depend on our mock catalog.
Specifically, for the images in our catalog, the number of subhalos in an image is drawn from a uniform distribution between 0 and 25 for the detectable mass bins ($m>10^{8}\msun$).
Similarly, we assumed a catalog of only bright sources with magnitudes less than 21.
To generalize this procedure to real lensing data, we would need to estimate the expected number of detectable subhalos per mass bin and the variance from $N$-body simulations.
As previously discussed, the ability of the network to detect substructure is determined by the brightness of the source, so the distribution of source brightness across real data will affect the extraction of the SMF slope.

%%%%%%%%%%%%%%%%%%%%%%%%%%%%%%%%%%%%%%%%%%%%%%%%%%%%%%%%%%%%%%%%%%%%%% 
\section{Discussion and outlook}
\label{sec:conclusion}
%%%%%%%%%%%%%%%%%%%%%%%%%%%%%%%%%%%%%%%%%%%%%%%%%%%%%%%%%%%%%%%%%%%%%% 

We developed a method to detect subhalos in strong gravitational lens images.
The method is based on image segmentation: we classify each pixel in an image as belonging to either the main lens, a subhalo within a given mass range, or neither.
We trained a convolutional neural network with a U-Net architecture on images with either no substructure or a single subhalo near the lensed images.
When the network is applied to an independent test set, it performs remarkably well.
In order for us to claim that a subhalo was correctly identified, the center of the subhalo can be at most 2 pixels off from the true center.
This corresponds to better than 0.12\arcsec~accuracy (given our pixel resolution), which would allow for good initial positions for follow-up studies with traditional methods.

We find that there are three common ways for the network to misclassify a pixel.
\begin{enumerate}
    \item The subhalo is not detected and all of the pixels are assigned to the background or main lens classes.
    This happens more for light subhalos than large subhalos.
    \item The pixel is on the edge of a subhalo and it is labeled as background instead of belonging to a subhalo. 
    In these cases, the network finds the subhalo, but it is predicted to be a few pixels too large or too small.
    \item Misidentifying the mass bin, generally by assigning the pixel to an adjacent mass bin.
    In these cases, the subhalo is still located correctly, but the mass is shifted up or down by a bin.
\end{enumerate}

While the network is trained specifically looking pixel by pixel, we need to cluster the detected pixels into subhalos to extract some physical meaning to the pixel-based accuracies.
On average, the subhalos have an area of $4\pi$ pixels by design, allowing for an easy conversion between the number of tagged pixels and the number of subhalos.
Pixels around the edge of a subhalo can be missed while allowing for a detection of the subhalo.
Because of this, the subhalo detection accuracy is better than the per-pixel accuracy.

With noise from a simulated single HST orbit, we are able to detect 17\% of the $10^{9}\msun$ subhalos in pixels near the Einstein ring, which are at least 50\% as bright as the brightest pixel for source brighter than magnitude 21.
The accuracy increases for brighter pixels and decreases for dimmer pixels.
The detection ability quickly increases for heavier subhalos; for instance, 45\% of the $10^{9.5}\msun$ subhalos are detected when located in the bright pixels.
We note that the accuracy also improves if the noise level can be reduced either through longer exposure or improved instrumentation.
As an example, the percentage of detected subhalos in the $10^{9}\msun$ bin rises from 17\% to 59\% when we reduce the noise by including 50 orbits of exposure.
This would also allow for the detection of even lighter subhalos.

To put this in perspective, \cite{2005MNRAS.363.1136K} and \cite{2009MNRAS.392..945V} showed that gravitational imaging can find subhalos with masses of a few times $10^8\msun$ for a signal-to-noise ratio of as low as 3 when the subhalo has an NFW profile and on the Einstein ring.
Additionally, these methods have detected subhalos with masses of $(3.51 \pm 0.15)\times10^9\msun$~\citep{2010MNRAS.408.1969V} and $(1.9\pm0.1)\times10^8\msun$~\citep{2012Natur.481..341V}.
These detections would fall into our $10^{9.5}\msun$ and $10^{8.5}\msun$ bins, respectively.
An advantage of our method is that we do not need to initially model the smooth lens to detect substructure in the system, which can take $\mathcal{O}$(weeks) to analyze in real systems with complex sources.
The modeling would take less time for sources modeled with a single Sersic, as done in this work.
It takes us less than a second to run an image through the network.
 
The network was also tested on images that do not contain substructure.
False substructure was predicted at a rate of around 3 subhalos per 100 images.
Most of these fall into the lightest mass bins that a network is sensitive to.
The good detection accuracy and low false-positive rate implies that if the network predicts substructure is present an image, it is mostly likely truly due to a subhalo being present. 

We also applied the network trained on images with single subhalos to images with many subhalos.
We showed that the network has out-of-sample adaptability and can generalize to identify an abundance of substructure in a single image, although it does not see any such data during training.
Traditional techniques would need to evaluate and fit models with different fixed number of subhalos, whereas the network makes the predictions automatically.
With this, we examined a method to determine the SMF power law index from the network output of multiple images.
The SMF is a key target for dark matter science because it can diagnose deviations from the standard cold dark matter scenario.

The technique uses a likelihood ratio for the count of detected subhalos in all of the bins, taking into account the expected counts given the power law index and the network's detection accuracy for each class.
We estimate that a $10\%$ determination of the slope of the SMF will require around 1000 images for images with one HST orbit.
However, this is strongly dependent on astrophysical assumptions concerning the total mass of a halo contained in substructure.
For instance, our catalog contained images with [0, 25] subhalos with masses $m\geq10^8\msun$.
If the true distribution of subhalos produces significantly more or less subhalos near the Einstein ring, the number of images necessary for an accurate determination of the power law index of the SMF could change.

We also note that \cite{2019ApJ...886...49B} uses likelihood-free inference to extract population-level substructure properties about the dark matter subhalo population.
They use a neural network to approximate the likelihood ratio function to obtain both the slope and the amplitude of the SMF with $\mathcal{O} (100)$ images.
In contrast, we infer about the SMF by explicitly resolving individual subhalo, which can then be further studied.
 
In the near future, the Vera Rubin Observatory is expected to find $10^4$ galaxy--galaxy strong lenses~\citep{2009arXiv0912.0201L} and Euclid expects $10^5$ galaxies lensed by galaxies in the field-of-view~\citep{2010arXiv1001.0061R}.
The ability to quickly detect substructure in these lenses could dramatically improve our understanding of dark matter.
Image segmentation is a promising method to study dark matter subhalos.
We stress that work will be needed to increase the robustness to noisier images (dimmer sources).
Additionally, we have ignored the possibility of extra perturbations to the lens along the line of sight \citep{D_Aloisio_2010,Li:2016afu,McCully:2016yfe,Despali:2017ksx,Sengul:2020yya}, but not part of the main lens halo.
One potential shortcoming of machine-learning methods is their ability to handle complex light sources.
In this, and other machine-learning subhalo detection studies, the light source is modeled with a single Sersic profile.
More study is needed to improve their robustness to realistic sources.

%%%%%%%%%%%%%%%%%%%%%%%%%%%%%%%%%%%%%%%%%%%%%%%%%%%%%%%%%%%%%%%%%%%%%% 
\begin{acknowledgments}
We thank Simon Birrer, A. Cagan Sengul, and Arthur Tsang for help with \textsc{Lenstronomy}.
We are thankful to Simon Birrer, Johann Brehmer, and Siddharth Mishra-Sharma for insightful comments on a previous version of this paper.
B.O. was supported in part by the U.S. Department of Energy under contract DE-SC0013607.
C..D was partially supported by the Department of Energy (DOE) grant No. DE-SC0020223.
The computations in this paper were run on the FASRC Cannon cluster supported by the FAS Division of Science Research Computing Group at Harvard University.
\end{acknowledgments}

\appendix

\setcounter{equation}{0}
\setcounter{figure}{0}
\setcounter{section}{0}
\setcounter{table}{0}
\makeatletter

\renewcommand{\theequation}{A\arabic{equation}}
\renewcommand{\thefigure}{A\arabic{figure}}
\renewcommand{\thetable}{A\arabic{table}}

%%%%%%%%%%%%%%%%%%%%%%%%%%%%%%%%%%%%%%%%%%%%%%%%%%%%%%%%%%%%%%%%%%%%%% 
\section{Single subhalo pixel predictions}
\label{sec:appSingleSubhalo}

\begin{figure}[h]
    \centering
    \includegraphics[width=0.75\linewidth]{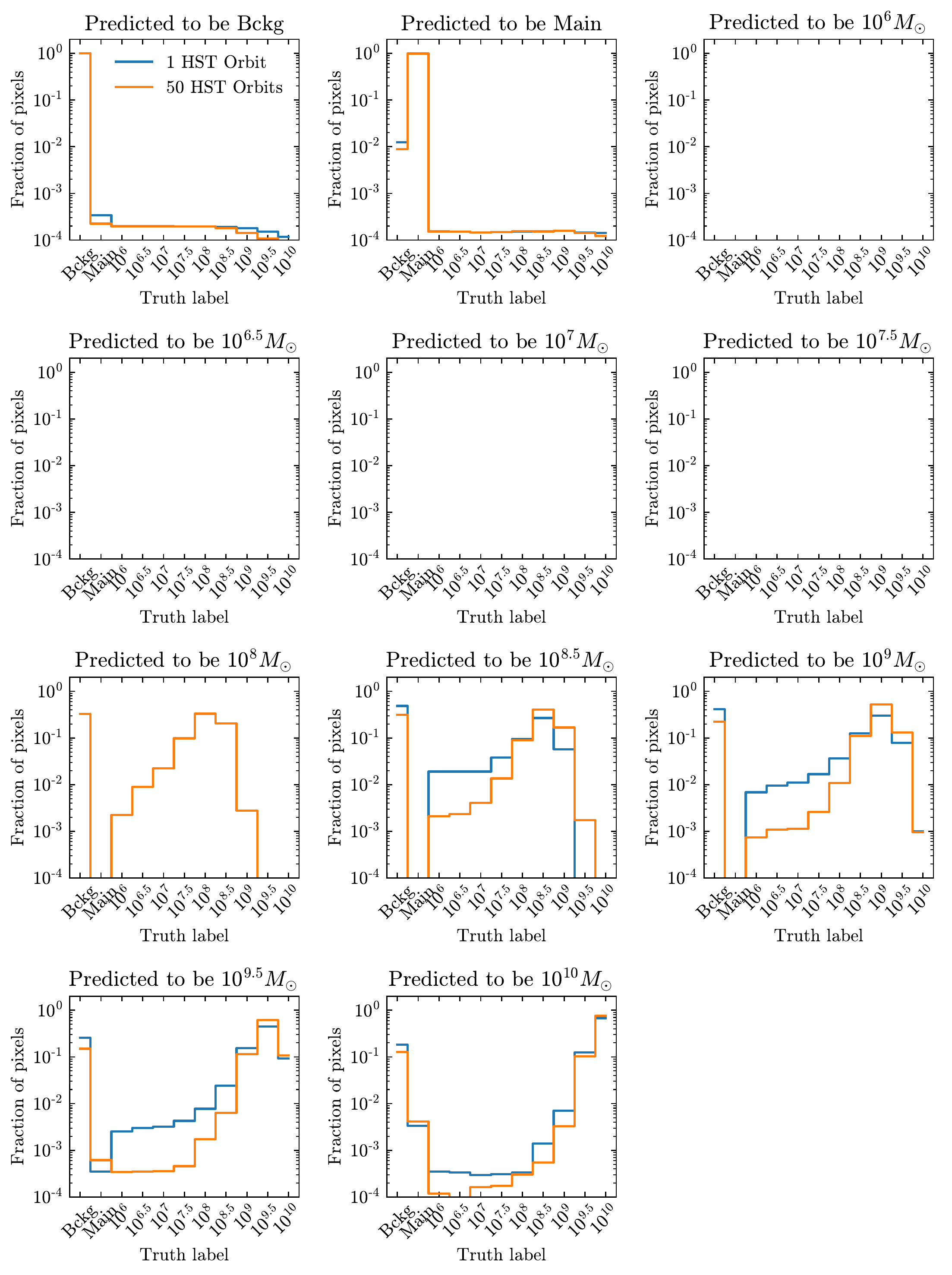}
    \caption{
    Each panel corresponds to pixels which are predicted of the indicated class.
    The $x$-axis denotes the class that the pixels belong to at truth level.
    Each panel is normalized to unity. 
    The results for images with one and 50 orbits of exposure are shown in the blue and orange lines, respectively.
    }
    \label{fig:TrueVPred}
\end{figure}

In the main text, we examine the per-pixel predictions on the images of the test set in Fig.~\ref{fig:PredVTrue}.
There, each panel corresponded to the true label of the pixel and the $x$-axis represented the predicted class for the pixel.
Here, we flip the information around.
Each panel in Fig.~\ref{fig:TrueVPred} corresponds to pixels that are predicted to be part of the class indicated by the title.
The $x$-axis then shows the true label and the panels are again normalized to unity.
The blue and orange lines are for images with one and 50 orbits of exposure, respectively.
It is now clear that the class that is predicted is very likely to be correct.
The networks never predict the lightest subhalo mass bins, because they cannot detect them.
When the network does predict a subhalo, the most probable bin is the correct one.

%%%%%%%%%%%%%%%%%%%%%%%%%%%%%%%%%%%%%%%%%%%%%%%%%%%%%%%%%%%%%%%%%%%%%% 
\bibliography{segmentation}
\bibliographystyle{aasjournal}
%%%%%%%%%%%%%%%%%%%%%%%%%%%%%%%%%%%%%%%%%%%%%%%%%%%%%%%%%%%%%%%%%%%%%% 

\end{document}